\documentclass[twocolumn]{aastex63}

\received{May 3, 2021}
\revised{June 16, 2021}
\accepted{July 1, 2021}

\shorttitle{Characteristic scales of magnetic switchback patches near Sun}
\shortauthors{Fargette et al}
\graphicspath{{./}{figures/}}

\usepackage{listings}
\usepackage{amsmath}
\usepackage{amsfonts}
\usepackage{amssymb}
\usepackage{hyperref}
\usepackage{ulem}
\usepackage{verbatim}
\begin{document}

\title{Characteristic scales of magnetic switchback patches near Sun and their possible association with solar supergranulation and granulation}

\correspondingauthor{Naïs Fargette}

\author[0000-0001-6308-1715]{Naïs Fargette}
\affiliation{Institut de Recherche en Astrophysique et Planétologie, CNRS, UPS, CNES, Université de Toulouse, Toulouse, France}

\author[0000-0001-6807-8494]{Benoit Lavraud}
\affiliation{Institut de Recherche en Astrophysique et Planétologie, CNRS, UPS, CNES, Université de Toulouse, Toulouse, France}
\affiliation{Laboratoire d'astrophysique de Bordeaux, Univ. Bordeaux, CNRS, Pessac, France }

\author[0000-0003-4039-5767]{Alexis Rouillard}
\affiliation{Institut de Recherche en Astrophysique et Planétologie, CNRS, UPS, CNES, Université de Toulouse, Toulouse, France}

\author[0000-0002-2916-3837]{Victor Réville}
\affiliation{Institut de Recherche en Astrophysique et Planétologie, CNRS, UPS, CNES, Université de Toulouse, Toulouse, France}

\author[0000-0002-4401-0943]{Thierry Dudok De Wit}
\affiliation{LPC2E, CNRS, University of Orléans, CNES, Orléans, France}

\author[0000-0001-5315-2890]{Clara Froment}
\affiliation{LPC2E, CNRS, University of Orléans, CNES, Orléans, France}

\author[0000-0001-5258-6128]{Jasper S. Halekas}
\affiliation{Department of Physics and Astronomy, University of Iowa, Iowa City, Iowa, United States}

\author[0000-0002-6924-9408]{Tai Phan}
\affiliation{Space Sciences Laboratory, University of California, Berkeley, Berkeley, CA, USA }

\author[0000-0003-1191-1558]{David Malaspina}
\affiliation{ Department of Astrophysical and Planetary Sciences, University of Colorado, Boulder, CO, USA }
\affiliation{Laboratory for Atmospheric and Space Physics, University of Colorado, Boulder, CO, USA }

\author[0000-0002-1989-3596]{Stuart D. Bale}
\affiliation{Space Sciences Laboratory, University of California, Berkeley, Berkeley, CA, USA }
  
\author[0000-0002-7077-930X]{Justin Kasper}
\affiliation{Climate and Space Sciences and Engineering, University of Michigan, Ann Arbor, MI, US }

\author[0000-0003-2783-0808]{Philippe Louarn}
\affiliation{Institut de Recherche en Astrophysique et Planétologie, CNRS, UPS, CNES, Université de Toulouse, Toulouse, France}

\author[0000-0002-3520-4041]{Anthony W. Case}
\affiliation{Smithsonian Astrophysical Observatory, Cambridge, Massachusetts, US }

\author[0000-0001-6095-2490]{Kelly E. Korreck}
\affiliation{Smithsonian Astrophysical Observatory, Cambridge, Massachusetts, US }

\author[0000-0001-5030-6030]{Davin E.Larson}
\affiliation{Space Sciences Laboratory, University of California, Berkeley, Berkeley, CA, USA }

\author[0000-0002-1573-7457]{Marc Pulupa}
\affiliation{Space Sciences Laboratory, University of California, Berkeley, Berkeley, CA, USA }

\author[0000-0002-7728-0085]{Michael L. Stevens}
\affiliation{Smithsonian Astrophysical Observatory, Cambridge, Massachusetts, US }

\author[0000-0002-7287-5098]{Phyllis L.Whittlesey}
\affiliation{Space Sciences Laboratory, University of California, Berkeley, Berkeley, CA, USA }

\author[0000-0001-6235-5382]{Matthieu Berthomier}
\affiliation{Laboratoire de Physique des Plasmas, CNRS, Sorbonne Universite, Ecole Polytechnique, Observatoire de Paris, Université Paris-Saclay, Paris, France }


\begin{abstract}

Parker Solar Probe (PSP) data recorded within a heliocentric radial distance of 0.3~AU have revealed a magnetic field dominated by Alfvénic structures that undergo large local variations or even reversals of the radial magnetic field. They are called magnetic switchbacks, they are consistent with folds in magnetic field lines within a same magnetic sector, and are associated with velocity spikes during an otherwise calmer background. They are thought to originate either in the low solar atmosphere through magnetic reconnection processes, or result from the evolution of turbulence or velocity shears in the expanding solar wind. In this work, we investigate the temporal and spatial characteristic scales of magnetic switchback patches. We define switchbacks as a deviation from the nominal Parker spiral direction and detect them automatically for PSP encounters 1, 2, 4 and 5. We focus in particular on a 5.1-day interval dominated by switchbacks during E5. We perform a wavelet transform of the solid angle between the magnetic field and the Parker spiral and find periodic spatial modulations with two distinct wavelengths, respectively consistent with solar granulation and supergranulation scales.  In addition we find that switchback occurrence and spectral properties seem to depend on the source region of the solar wind rather than on the radial distance of PSP. These results suggest that switchbacks are formed in the low corona and modulated by the solar surface convection pattern.
\end{abstract}



\section{Introduction} \label{sec:1_intro}

From the very first encounter of Parker Solar Probe (PSP) with the Sun, in-situ data showed striking unexpected features: the solar wind was pervaded with frequent magnetic deflections, Alfvénic in nature, often showing velocity spikes and large radial magnetic field changes (\cite{2019Natur.576..228K},  \cite{2019Natur.576..237B}, \cite{2020ApJS..246...34P}).
Because they often show reversals of the radial magnetic field, they are usually called magnetic switchbacks. They are interpreted as localized folds of the magnetic field, considering that the supra-thermal electron strahl mostly remains unchanged throughout the structures \citep{2019Natur.576..228K}. This interpretation is also supported by observations of the differential streaming of alpha particles \citep{yamauchi_2004} and proton beams \citep{Neugebauer_2013}, and also confirmed by PSP (e.g. \cite{Woolley_2020}). Their association with velocity spikes stems from their Alfvénic nature, so that the spherical polarisation of the fluctuations and constant magnetic field intensity during these periods, as for example shown by \cite{Matteini_2014}. Their omnipresence in PSP data makes them a key feature in the inner heliosphere. The fact that they carry kinetic energy means that they may be related to the solar wind acceleration and the heating of the corona.

Albeit drawing less attention, these structures have been observed in the solar wind in past mission data. \cite{1999GeoRL..26..631B} noticed magnetic field polarity inversions at high heliographic latitudes with the Ulysses spacecraft, and interpreted them as large-scale folds in the magnetic field. \cite{2011ApJ...737L..35G} identified Alfvénic structures in WIND data during slow solar wind intervals of 320 to 550~km/s at 1~AU, stating that they may stem from Alfvénic turbulence. \cite{2018MNRAS.478.1980H} more recently analyzed such velocity spikes at 0.3 AU in fast streams (700~km/s) and conjectured that they could be signatures of transient events in the chromosphere or corona. It is now clear that at least part of the switchbacks are formed below PSP's orbit, either in the low atmosphere, or in between the Sun and the spacecraft. In the first case they would be remnants of solar wind formation, and in the second case a byproduct of solar wind evolution.

Several processes at the Sun's surface could explain such magnetic field deflections. \cite{2020ApJ...894L...4F} propose that interchange reconnection (\cite{2001ApJ...560..425F}, \cite{2005ApJ...626..563F}) can be a driver of switchback formation. Magnetic reconnection between open and closed field lines in the low atmosphere would generate folds in the magnetic field, and the latter would propagate into the interplanetary medium. Simulations by \cite{2020ApJS..246...32T} support that such a magnetic fold, once formed, would indeed survive long enough to be observed by PSP. In an alternate view proposed by \cite{2021AA_DRAKE}, interchange reconnection would rather produce flux ropes in the low corona. The flux ropes are advected by the solar wind and generate the reversal signatures observed by PSP. 
Finally, \cite{Schwadron_2021} have proposed that magnetic footpoints moving into a coronal hole (through interchange reconnection) would create a skewed magnetic connection between slow and fast wind called the super Parker spiral. The radial component of such a configuration reverses as it propagates through the corona, thus creating a switchback structure.

But one has to keep in mind that for the most part switchbacks are strikingly highly Alfvénic (\cite{2019Natur.576..237B}, \cite{2020ApJS..246...34P}). This points directly to Alfvénic turbulence as a source for the phenomenon, and \cite{Squire_2020} reproduce magnetic field reversals stemming from Alfvénic turbulence in 3D simulations, while \cite{Ruffolo_2020} show that non-linear shear driven turbulence can also create switchbacks.

The origin of these magnetic structures thus remains poorly understood, and their formation process is highly debated. Statistical studies have been performed on switchbacks, finding that they show no obvious characteristic duration and that their magnetic field power spectral density differs from the pristine solar wind \citep{2020_DDW}. Switchbacks tend to agregate in "patches", meaning that their occurrence is modulated at large scales (\cite{2019Natur.576..237B}, \cite{2020_DDW}), and their duration distribution is also found to be consistent with high aspect ratio structures (\cite{2020ApJS..246...45H}, \cite{2021AA_LAKER}).

One main idea discussed in this paper is that if switchbacks are formed in the low solar atmosphere as proposed by \cite{2020ApJ...894L...4F}, \cite{2021AA_DRAKE} or \cite{Schwadron_2021}, then they are probably affected - if not caused - by physical phenomena that impact the Sun's low atmosphere structure. This includes \citep{MEYER_2007_book} structures related to active regions (coronal loops, prominences, that may erupt into flares or coronal mass ejections), coronal bright points and plumes, spicules, as well as the convective motions at the Sun's surface observed as granulation \citep{2009LRSP....6....2N} and supergranulation \citep{2010LRSP....7....2R}. The latter are indeed believed to play a role in heating and accelerating the solar wind, as surface convection generates Alfvén waves that propagate along magnetic field lines, and some dissipate in the higher corona through turbulence (e.g. \cite{Velli_1989}, \cite{2007ApJS..171..520C}, \cite{2016ApJ...821..106V}).
Interchange reconnection as proposed by \cite{2020ApJ...894L...4F} would also occur at or near the supergranular network, as indeed the closed and open field lines involved should have magnetic footpoints rooted in the network, which in turn outlines the boundaries of supergranules (\cite{2009A&A...495..945R}, \cite{2010LRSP....7....2R}). Finally in earlier work on Helios Data, \cite{Thieme_1989} found structures in the solar wind density and velocity that were consistent in angular size with supergranulation, suggesting that its signature can be detected in the solar wind up to 0.7 AU.

In this work we investigate typical temporal and spatial scales associated with the switchback phenomenon through wavelet transforms, that could hint to a specific formation process. Our results mainly concern the in-situ modulation (patches) of switchback occurrence, hence corresponding to a larger scale phenomenon than individual magnetic switchbacks. In section \ref{sec:2_method} we present the different data products and detail the process of switchback identification as well as the spatial projection we perform. In section \ref{sec:3_enc5} we focus on encounter 5 and perform both temporal and spatial wavelet analyses on switchbacks for a 5.1-day interval. In section \ref{sec:4_enc2} we present the spatial analysis of encounter 2 in a similar manner. Finally in section \ref{sec:5_disc} we discuss scales associated with potential formation processes, in particular those related to solar wind turbulence and solar convection patterns.
\section{Methods} 
\label{sec:2_method}

\subsection{Data analysis}
\label{sec:2.1_data}
We analyze magnetic field data provided by the FIELDS instrument suite \citep{2016SSRv..204...49B} and particle data provided by the Solar Wind Electrons Alphas and Protons (SWEAP) instrument suite \citep{2016SSRv..204..131K}. The latter includes plasma moments from the Solar Probe Cup (SPC) \citep{2020ApJS..246...43C} and electron pitch angle distributions from the Solar Probe ANalyzers (SPANs) \citep{2020ApJS..246...74W}. Data is shown in the RTN frame of reference, with \textbf{R} the Sun to spacecraft unit vector, \textbf{T} the cross product between the Sun's spin axis and \textbf{R}, while \textbf{N} completes the direct orthogonal frame. In this work we focus on data taken by PSP below 60 Solar Radii ($R_s$) during encounters 1, 2, 4 and 5 (thereafter noted $E_x$) with an emphasis on $E_5$ and $E_2$. We do not consider $E_3$ as SPC data are not available for most of the encounter. PSP went down to 36~$R_s$ during $E_1$ and $E_2$, and down to  28~$R_s$ during $E_4$ and $E_5$.

In order to discard high frequency kinetic effects, as well as to reduce instrumental noise, we re-sample all data products from SPC and FIELDS to a constant time step taken at 2~seconds (\cite{2020_DDW}). The sampling is done by using a 1 dimensional B-spline interpolation, a method available through the scipy.interpolate package in Python (\cite{1993csfw.book.....D}).

\subsection{Switchback definition and identification}
\label{sec:2.2_SB}
To identify switchbacks in a systematic manner, we define them as a deviation from the Parker spiral, as done by \cite{2020ApJS..246...45H}. The Parker spiral angle $\psi_p(t)$ is the trigonometric angle between the radial direction and the spiral direction in the RTN plane, given by 
\begin{equation}
    \psi_p(t) = \arctan \left( \dfrac{-\omega \left(r(t)-r_0\right)} {V_r(t)}\right) + k \pi
\end{equation} \\
where $\omega = 2.9\times10^{-6}$~s$^{-1}$ is the Sun's angular moment taken at the equator, $r(t)$ is the distance of the spacecraft to the center of the Sun, $r_0=1$~$R_S$ is the source of the Parker spiral, $k$ is a dimensionless integer equal to 0 (anti-sunward field) or 1 (sunward field), and $V_r(t)$ is the measured radial speed of the solar wind averaged over two hours. This averaging allows for the removal of short timescale variations and transient structures that are not relevant to the Parker spiral angle. We then reconstruct a semi-empirical vector for the Parker spiral magnetic field $\mathbf{B_{p}}(t)$, contained in the RT plane while keeping the field amplitude measured by PSP $|\mathbf{B}(t)|$: 
\begin{equation}
    \mathbf{B_p}(t) =|\mathbf{B}(t)| \begin{bmatrix} \cos{\psi_p(t)}\\ \sin{\psi_p(t)}\\ 0
\end{bmatrix}_{RTN}
\end{equation}
The normalized solid angle $\widetilde{ \Omega}$ between $\mathbf{B_p}$ and $\mathbf{B}$ \citep{2020_DDW} is then given by 
\begin{equation}
    \widetilde{ \Omega}(t) = \dfrac{1}{2} \left(1 - \cos{\alpha (t)}\right)
\end{equation}
with
\begin{equation}
    \cos{\alpha(t)} =  \dfrac{\mathbf{B_p}(t). \mathbf{B}(t)}{ |\mathbf{B}(t)|^2}
\end{equation}
$\widetilde{ \Omega}$ reflects whether both vectors are aligned ($\widetilde{\Omega}=0$) or diametrically opposed ($\widetilde{\Omega}=1$). Switchbacks can then be detected automatically by setting a threshold on $\widetilde{ \Omega}$. This threshold will necessarily impact our results, and it has been taken in the literature at $\widetilde{ \Omega}=0.15$ ($\alpha = 45^o$) \citep{MacNeil_2020}, or $\widetilde{ \Omega}=0.07$ ($\alpha = 30^o$) \citep{2020ApJS..246...45H}. One may also take $\widetilde{ \Omega}=0.5$ ($\alpha = 90^o$) to be consistent with the very idea of a switchback (a reversal of the radial magnetic field component). We add an additional detection condition that five consecutive points are needed to detect a switchback, this means that our study can only address timescales longer than 10 seconds. This is motivated by the fact that wave activity is present within switchbacks, and may lead to several crossings of the threshold line within one switchback.

\begin{figure*}
    \centering
    \includegraphics[width=1\textwidth]{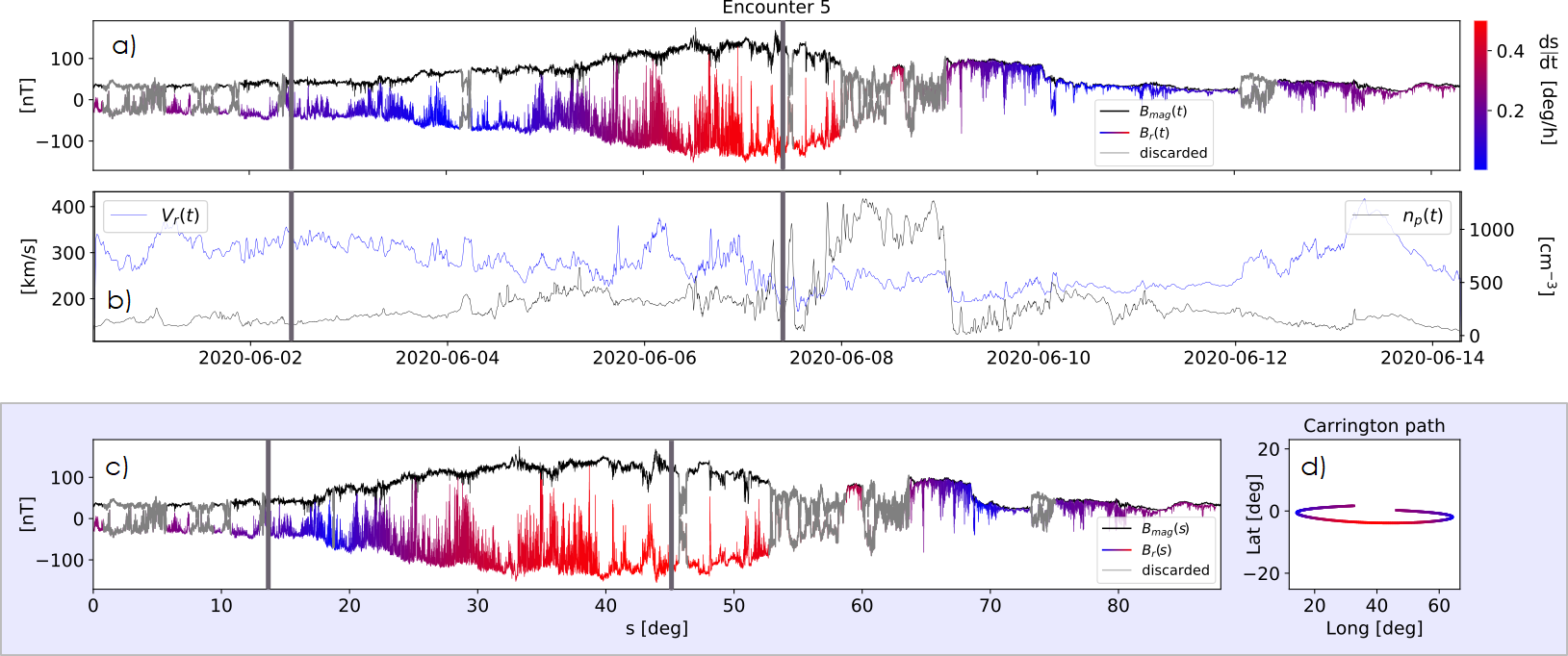}
    \caption{Encounter 5 context. a) Magnetic field amplitude $B_{mag}$ and radial component $B_R$, which is color coded by the spacecraft absolute velocity with respect to the solar surface, $ds/dt$ in degree/hour (see text for details). Grey data denote intervals that are discarded because they are irrelevant to the switchback study. Vertical grey lines indicates the region analysed. b) Solar wind radial velocity $V_r$ and ion density $n_p$ ; c)  $B_{mag}$ and $B_{r}$ are displayed in the same manner as in panel a) but now as a function of the angular distance $s$, d) PSP trajectory projected in Carrington coordinates and color coded with $ds/dt$. }
    \label{fig: 1_context}
\end{figure*}

The accuracy of this method depends on the adequacy of the Parker spiral model to represent the undisturbed magnetic field. Thus, in our study we discard intervals identified as heliospheric current sheet crossings and plasma sheets (see \cite{2020ApJS..246...47S} and  \cite{2020ApJ...894L..19L} for $E_1$), Magnetic Increases with Central Current Sheet (MICCS) structures (\cite{2021AA_FARGETTE}), coronal mass ejections (\cite{2020ApJS..246...63N}, \cite{2020ApJS..246...69K}), as well as periods of strahl drop out where magnetic field lines are most likely disconnected from the Sun \citep{Gosling_2006}. All of these intervals are identified visually while scanning through the data, and are given in the appendix in Table~\ref{tab:intervals}.

\subsection{Space time bijection}
\label{sec:2.3_s}

To study potential spatial scales associated with switchbacks, we need to transform the PSP time series into functions of a given spatial parameter. This might be achieved by different methods with varying degrees of complexity, for instance by taking the Carrington longitude of the spacecraft, or computing the Parker spiral footpoint coordinates, or even by calculating its connectivity coordinates with Potential Field Source Surface (PFSS) \citep{2020ApJS..246...23B} or magneto-hydrodynamics simulations \citep{2020ApJS..246...24R}. In this study, we chose not to use a ballistic projection of the Parker spiral on the Sun's surface, as it is poorly suited for spectral analysis. Indeed, when the radial velocity of the solar wind changes, the spiral footpoint can turn around, hence losing the one-to-one correspondence between time and space and folding the signal over itself. Instead, we decided to use a direct projection of the spacecraft path on the Sun's surface, using the angular displacement $s$ defined by:

\begin{equation}
    s(t) = \int_0^t{ds}
    \label{eq: 6}
\end{equation}
and 
\begin{equation}
    ds = \sqrt{d\theta^2 + \cos^2\theta~ d\phi^2}
    \label{eq: ds}
\end{equation}
where $\theta$ and $\phi$ are the Carrington latitude and longitude of the projected orbit over time. We also resample the data over a constant spatial step taken as $ds = \text{max}(s) / N_{point} $. This way we keep a similar number of measurement points $N_{point}$ between the spatial and temporal analysis. To convert $s$ to regular distances, one only needs to multiply it by the considered radius.

This choice of metric ensures a bijection between time and space, and takes into account the variation in both latitude and longitude. We note, however, that with this projection $s$ we make the assumption that when PSP remains within a given source area on the solar surface, the displacement of its footpoint is equivalent to the assumed displacement of the spacecraft projection. 
\section{Encounter 5} 
\label{sec:3_enc5}

\subsection{Context}
\label{sec:3.1_context}

In Figure \ref{fig: 1_context}, we display for context the magnetic field magnitude $B_{mag}$ and radial component $B_{R}$ through $E_5$ (\ref{fig: 1_context}a), as well as the solar wind radial velocity $V_{r}$ and proton density $n_{p}$, both averaged over 30~minutes (\ref{fig: 1_context}b). Grey data correspond to intervals that were discarded as detailed in section \ref{sec:2.2_SB} (Table \ref{tab:intervals}). $B_{mag}$ scales as $r^{-2}$ and reaches 137~nT at perihelion on 2020-06-07~08:20. The polarity is negative until the HCS crossing (from 2020-06-08~00:00 to 2020-06-09~01:40) and remains positive thereafter. The spacecraft is sampling slow solar wind below 420~km/s with an average value of 274~($\pm$~46)~km/s. On the other hand the density increases as expected during plasma sheet and HCS crossings, reaching up to 1200~cm$^{-3}$. Outside these intervals, $n_p$ scales as $r^{-2}$ and reaches around 400~cm$^{-3}~$ at perihelion.

\begin{figure*}
    \centering
    \includegraphics[width=1\textwidth]{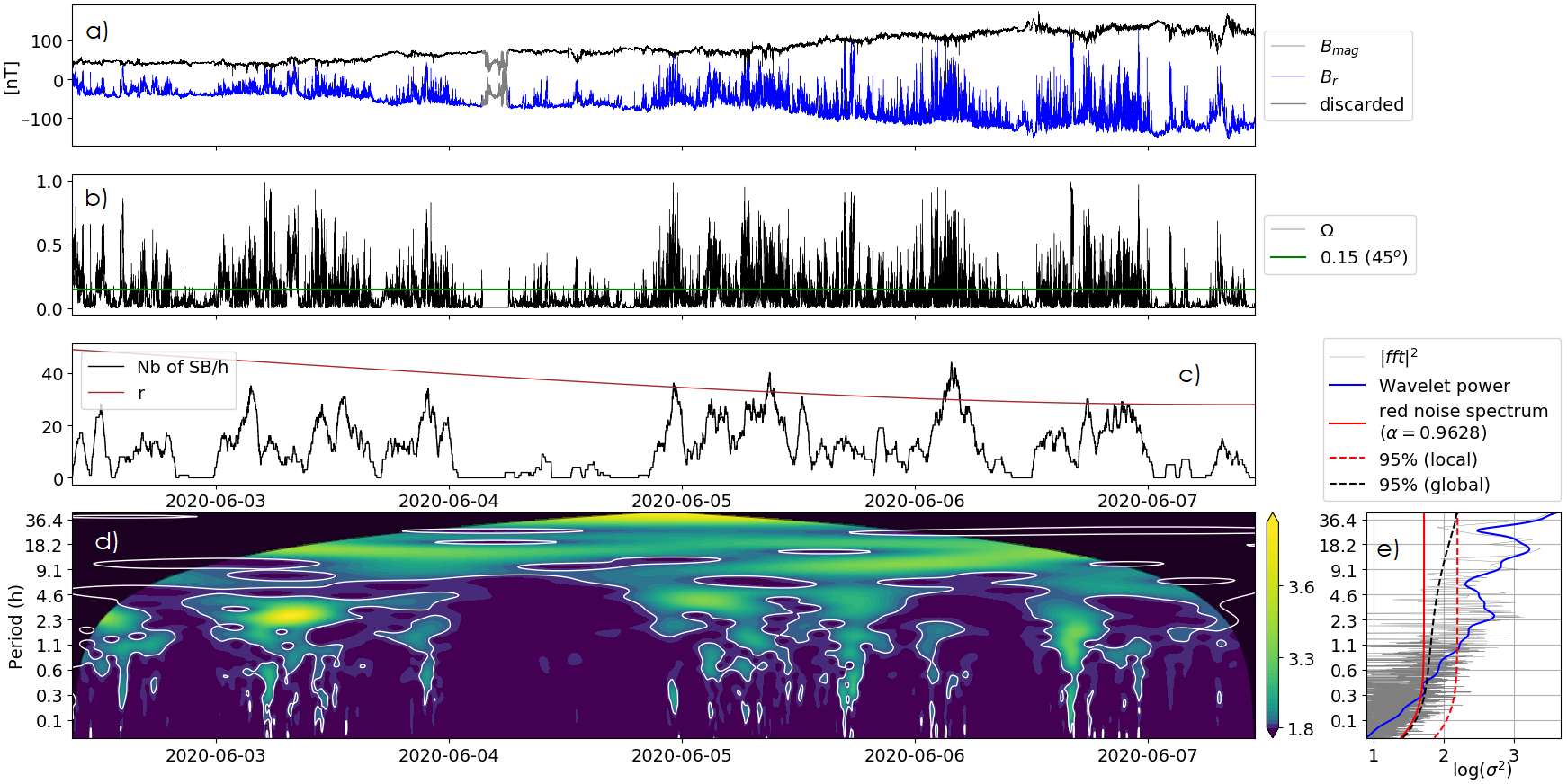}
    \caption{Temporal analysis for $E_5$ a) $B_{mag}$ and $B_R$ plotted as a function time b) the solid angle $\Omega$ with a horizontal green line indicating the switchback detection threshold set at 0.15. c) the number of switchback detected per hour together with $r$ the radial distance of PSP to the Sun. d) Wavelet Power Spectrum (WPS) of $\widetilde{\Omega}$ performed over periods of 0.08 to 42.6 hours and represented in a logarithmic scale. White contours represent the local 95\% confidence level e) FFT of $\widetilde{\Omega}$ in light gray, integrated WPS in blue, theoretical red noise spectrum in red, 95\% local confidence level in dashed red and global 95\% local confidence level in dashed black (units similar to panel d).}
    \label{fig: 2_time}
\end{figure*}

To illustrate the spatial projection we perform, $B_R$ in panel \ref{fig: 1_context}a is color coded with the absolute speed of the spacecraft relative to the Sun's surface, defined in equation \ref{eq: ds}. In Figure \ref{fig: 1_context}c,  $B_{mag}$ and $B_{r}$ are plotted relative to $s$ with the same color scale, and Figure \ref{fig: 1_context}d displays the path of PSP on the Sun's surface in Carrington coordinates. Logically, periods of co-rotation (in blue) are shortened in the spatial representation (\ref{fig: 1_context}c) compared to the temporal one (\ref{fig: 1_context}a). 

We now restrict our analysis to the period comprised between 2020-06-02~09:10 and 2020-06-07~11:00 (vertical lines in the figure). This interval is indeed characterized by persistent and stable patches of switchbacks, and the frequency analysis performed next requires a signal as continuous as possible. The succession of strahl drop outs and flux ropes before 2020-06-02~09:10 or the HCS crossing after perihelion would bias our analysis. We also consider in this case that the plasma sheet observed around 2020-06-04~04:30 is sufficiently small to be included in our signal. Overall we are studying 5.1~days of data, covering 32.1~degrees of angular distance with a constant spatial step of $ds = 1.5\times 10^{-4}$ degrees.

\label{sec: 3.2_time}

In Figure \ref{fig: 2_time}, we display the temporal analysis of switchbacks over the selected time period. Panel \ref{fig: 2_time}a recalls the magnetic field for clarity, panel \ref{fig: 2_time}b displays the solid angle to the Parker spiral $\widetilde{\Omega}$ together with a threshold detection at 0.15 (45$^o$), and panel \ref{fig: 2_time}c is the number of switchback per hour. We also over-plot $r$ the distance to the Sun in $R_s$. During the 2-hour partial heliospheric plasma sheet crossing (in grey in panel \ref{fig: 2_time}a) $\widetilde{ \Omega}$ is set to zero. 

\subsection{Radial dependence of switchback occurrence}

In panels \ref{fig: 2_time}a and \ref{fig: 2_time}b we see visible patches of switchbacks, as observed by \cite{2019Natur.576..228K} or \cite{2019Natur.576..237B} for instance. The number of switchbacks is on average 11~$h^{-1}$ over this whole 5.1-day period. Even though $r$ decreases from 50 to 28 $R_s$, the number of switchbacks does not seem to follow a conjugate decrease. During the patches occurring on June 3, $r$ decreases from 45.2 to 39.6~$R_s$ and the average switchback frequency is 14.8$\pm$~7.6~$h^{-1}$. On the other hand from June 5 to June 7 $r$ decreases from 34.4 to 28.0~$R_s$ and the average switchback frequency is of 15.2 $\pm$~9.0~$h^{-1}$. We also observe that following the plasma sheet observed on June 4, and preceding the one observed on June 7 (not shown, cf Figure \ref{fig: 1_context}), the number of switchbacks drops significantly below 5~$h^{-1}$on average. This suggests that the number of switchback detected with our method is  uncorrelated to the radial distance during this period, and is by contrast sensitive to the plasma environment and spacecraft connectivity.

\subsection{Temporal spectral analysis}
To further investigate the possible timescales associated with switchbacks we perform a wavelet analysis on the solid angle $\widetilde{\Omega}(t)$ based on \cite{1998BAMS...79...61T}. One difference with \cite{1998BAMS...79...61T} is that we do not detrend our data (i.e. we do not substract the low-frequency component to our signal) as advised by \cite{2016ApJ...825..110A} for instance. We use a Morlet wavelet as a mother wavelet, and all spectrum are hereafter normalized by $N_{point}/(2\sigma^2)$, $\sigma^2$ being the data variance. 
We display the wavelet power spectrum (WPS) of the signal in panel \ref{fig: 2_time}d with a logarithmic colormap, for periods from 5 minutes (150*$dt$) to 42.6 hours (one third of the considered period). We investigated below the 5 minute scale but found no distinct wavelength that stood out in the WPS. This is consistent with previous results showing that individual switchbacks seen by PSP do not display any preferential duration. The blackened area denotes the cone of influence where the WPS is affected by edge effects and is not relevant.
As done by \cite{1998BAMS...79...61T} we use a red noise model as a background spectrum based on the lag-1 autoregressive process, and find a correlation coefficient of $\alpha$=0.9628. In panel \ref{fig: 2_time}e the Fourier power spectrum of $\widetilde{\Omega}(t)$ is displayed in grey, and the global wavelet power spectrum (integrated over time) is plotted in blue. The red solid curve denotes the theoretical power spectrum of red noise, and the dashed red curve the 95\% confidence level for the local spectrum, which yields the white contour in panel \ref{fig: 2_time}d. The black dashed curve is the 95\% confidence level for the global spectrum.
\begin{figure*}
    \centering
    \includegraphics[width=1\textwidth]{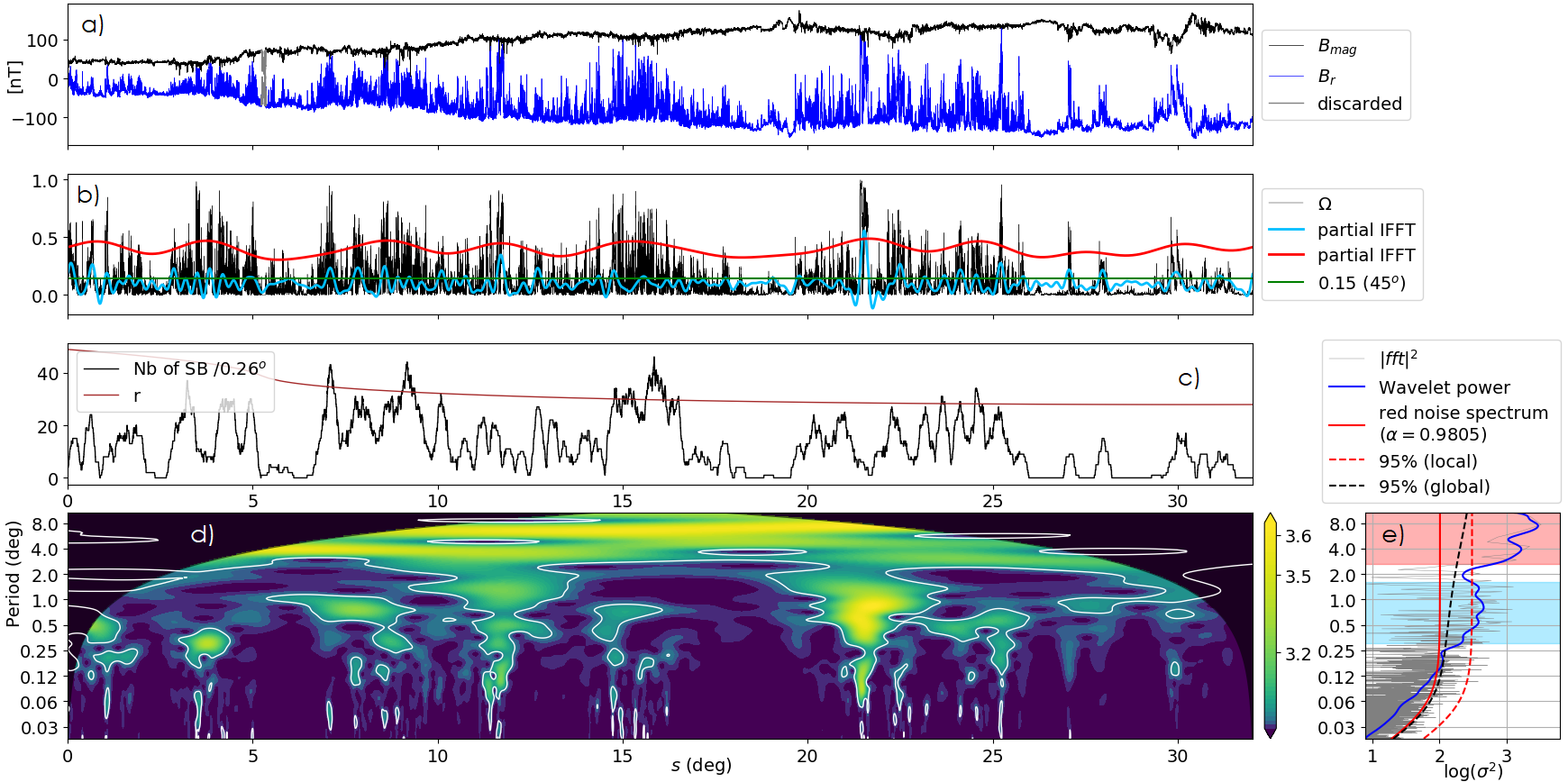}
    \caption{Spatial analysis for $E_5$ a) $B_{mag}$ and $B_R$ plotted as a function of the angular displacement $s$ b) the solid angle $\Omega$ with the horizontal green line indicating the switchback detection threshold set at 0.15 (45$^o$), and partial inverse FFT of  the peaks highlighted in panel e are over plotted c) the number of switchbacks detected per 0.26$^o$ together with $r$ the radial distance of PSP to the Sun. d) Wavelet Power Spectrum (WPS) of $\Omega$ performed over periods of 0.02 to 10.7 degrees and represented in a logarithmic scale. White contours represent the local 95\% confidence level e) FFT of $\Omega$ in light gray, integrated WPS in blue, theoretical red noise spectrum in red, 95\% confidence level to the local spectrum in dashed red and 95\% confidence level to the global spectrum in dashed black, while peaks in the FFT are highlighted in light red and light blue.}
    \label{fig: 3_space}
\end{figure*}

Figure \ref{fig: 2_time}e shows that several timescales are detected through this interval. At large scales, the WPS first peaks at a period between 13h and 18h, these periods correspond to the duration of the three large patches of switchbacks visible by eye in panel \ref{fig: 2_time}b on June 3. Then from June 5, the large scale period is less well defined but increases from 8h to 18h by the end of June 6. At shorter periods, the most visible feature occurs on June 3 where a large switchback dominates the spectrum and a periodicity of 2 to 5 hour is present, producing a broad peak in the global wavelet power (\ref{fig: 2_time}e). This wavelength persists on the beginning of June 5 and there corresponds to the duration of small patches of switchbacks. Overall, some significant wavelengths arise locally throughout the 5.1 day interval, but they are not particularly coherent or well organized. We nevertheless observe that patches of switchbacks last from 5 to 18 hours. The next section supports that when analyzed spatially the signatures are more consistent.
\begin{figure*}
    \centering
    \includegraphics[width=1\textwidth]{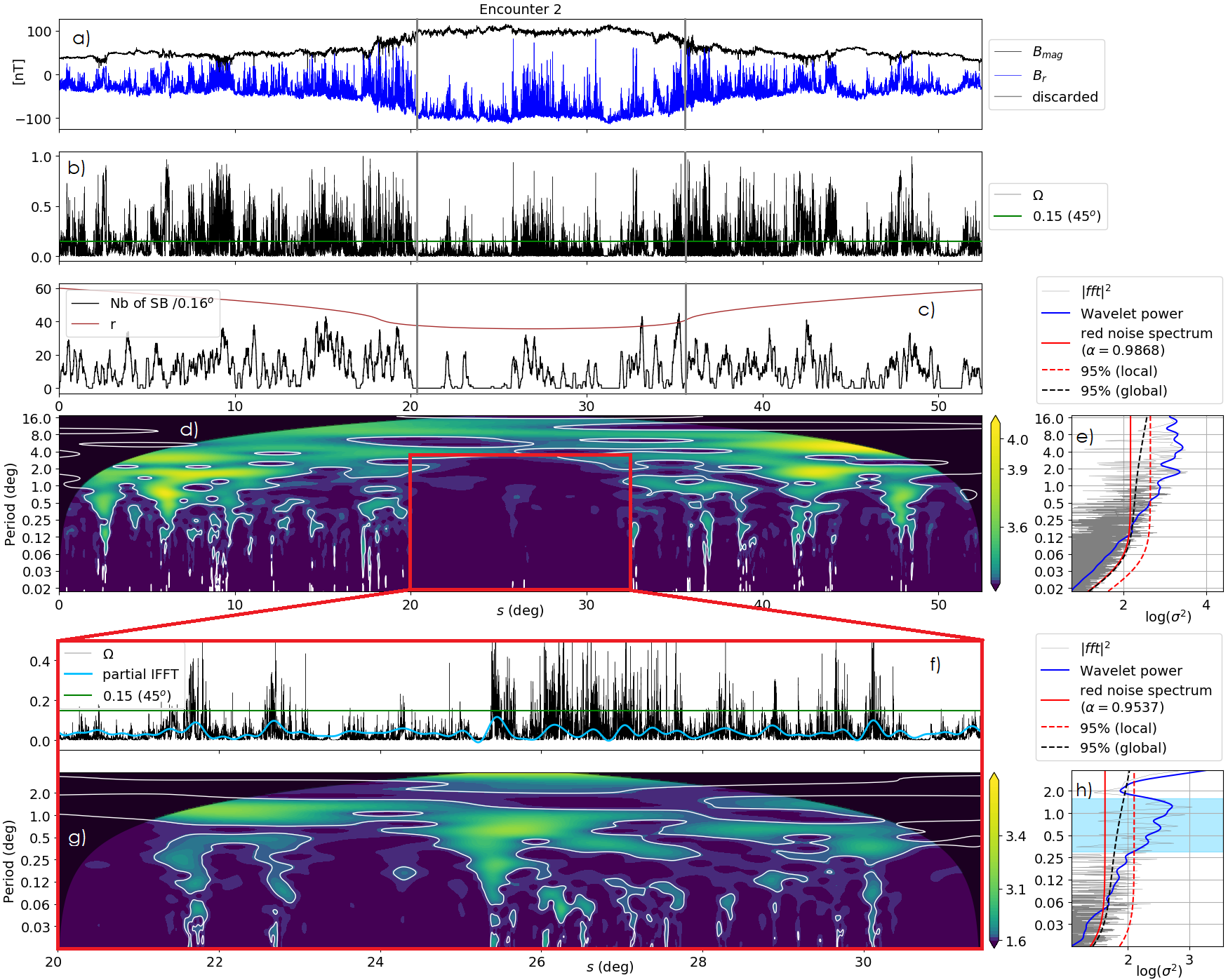}
    \caption{Encounter 2 spatial analysis. We plot the parameters in the same manner as in Figure \ref{fig: 3_space} for panels a) to e). The vertical grey lines at $s$ = 20.3$^o$ and $s$ = 35.6$^o$ denotes the change in plasma properties between streamer and coronal hole plasmas (see text for further details). In panels f) to h) we present a zoom-in on the red rectangle indicated in panel d). We display in panel f) the solid angle $\widetilde{\Omega}$ with a green horizontal line for switchback threshold detection. The light blue curve represents the inverse Fourier transform of the peak visible in panel h) between 0.3$^o$ and 1.6$^o$ shaded in light blue. In panel g) and h) we display the WPS and Fourier transform in the same manner as in panels d) and e) }
    \label{fig: 4_E2}
\end{figure*}

\subsection{Spatial analysis}
\label{sec:3.3_space}
To identify potential spatial scales associated with magnetic switchbacks, we repeat the analysis of section \ref{sec: 3.2_time} but as a function of $s$ (see section \ref{sec:2.3_s}). We display the results in Figure~\ref{fig: 3_space} in the same manner as Figure~\ref{fig: 2_time}. In panel \ref{fig: 3_space}c we plot the number of SB for a spatial window of 0.26 degrees, this value is consistent with the 1h scale shown in Figure \ref{fig: 2_time}. The spectral analysis is performed on scales from 0.02 deg (150*$ds$) to 10.7 deg (one third of the considered interval). Regarding the red noise model we find a correlation coefficient of $\alpha$=0.9804.

A striking feature in the number of switchbacks in panel \ref{fig: 3_space}c is that marked periodicities arise in the signal, most visible between $11^o < s < 13.5^o$  and  $22^o <s<25.5^o$ (wavelength of 0.5$^o$), and $3^o<s<5^o$, $6^o<s<10^o$ and $20^o<s<22^o$ (wavelength of 1$^o$). Of course this observation depends strongly on the scale of 0.263$^o$ chosen here. Since we did not include a hysteresis in our detection, this regularity can be attributed to fluctuations of $\mathbf{B}$ around the chosen threshold of $\widetilde{\Omega} = 0.15$. Nonetheless it is the signature of a possible periodicity that we investigate further through the wavelet transform of $\widetilde{\Omega}(s)$ (\ref{fig: 3_space}d). 

Overall the WPS over $s$ highlights spatial scales that were not clearly present in the temporal analysis. First we can see that the three main patches of switchbacks visible to the eye in panel \ref{fig: 3_space}(b) from $s=5^o$ to $s=27^o$ have the same scale in order of magnitude. This is quantified by the WPS reaching its maximum consistently through the spatial series (\ref{fig: 3_space}d) and coincident with a peak in the integrated WPS (\ref{fig: 3_space}e) between periods of 2.6$^o$ and 10.7$^o$. Moreover, significant power is found at scales comprised between 0.3$^o$ and 1.6$^o$, particularly for $6^o<s<10^o$ (wavelength of 0.8$^o$) and for $22^o<s<25.5^o$  (wavelength of 0.5$^o$). This is consistent with the periodicity previously observed in panel \ref{fig: 3_space}c. The two peaks detected are broad, separated by one order of magnitude, and not always coincidental, meaning that the higher frequency one is unlikely to be a harmonic.

To further stress these wavelengths in the signal, we overplot in panel \ref{fig: 3_space}b the inverse of the signal's truncated Fourier transform, selecting only the peaks located between wavelength 0.3$^o$ and 1.6$^o$ (shaded in blue in panel \ref{fig: 3_space}e) and 2.6$^o$ and 10.7$^o$ (shaded in red in panel \ref{fig: 3_space}e). These partial inverse Fourier transform (IFFT) are translated upward in the panel by a constant value for clarity. They follow nicely the solid angle fluctuations for large (red) and meso-scale (light blue) patches. Finally, the analysis of the complete fifth encounter is available in the appendix, and we note that after the HCS crossing, the WPS once more highlights a persistent periodicity between 0.3$^o$ and 1.6$^o$.

What is remarkable in these spectral features is that the detected periodicity lasts for several wavelengths and they are moreover consistently observed throughout $E_5$. \ These results indicate that significant periodicity may arise locally in the magnetic field fluctuations. Comparison of these scales to physical phenomena are discussed in section \ref{sec:5_disc}.

\section{Encounter 2} 
\label{sec:4_enc2}

In this section we highlight some interesting features of the spatial analysis performed for $E_2$. This flyby of the Sun is particularly interesting because for one the observation of switchbacks is not interrupted by HCS crossings, CMEs or too frequent strahl dropouts, and in addition the spacecraft samples two different types of solar wind (\cite{2020ApJS..246...37R}, \cite{Griton_2021}). Until April 3, 2019 09:00 UT and from April 7, 2019 18:00 UT, PSP is sampling a high density slow solar wind that \cite{2020ApJS..246...37R} associate with streamer belt plasma through a white light imaging analysis. In between these dates it scans a lower density solar wind more probably associated with a coronal hole. 

In Figure \ref{fig: 4_E2} we display the spatial analysis for $E_2$ as in Figure \ref{fig: 3_space} (panels a to e). This represents 15 days of data and 52$^o$ covered. Vertical gray lines separate the regions identified by \cite{2020ApJS..246...37R} in panels a, b and c. In panel \ref{fig: 4_E2}c we observe that the number of switchbacks decreases in the coronal hole around perihelion. In streamer belt plasma, no obvious trend is visible while $r$ vary significantly. This is consistent with section \ref{sec:3_enc5}, where we find that switchback occurrence is sensitive to the plasma environment. 
Furthermore, it is obvious in panel \ref{fig: 4_E2}d that spectral properties are different between the different types of plasma. In both intervals of streamer belt plasma, two peaks are detected  with scales of respectively 2$^o$ and 5$^o$. By contrast in the coronal hole, no significant structures is visible below 4$^o$. This strongly suggests that the fluctuation properties differ with the solar wind source.

We renew the analysis on the area highlighted with a red rectangle in panel \ref{fig: 4_E2}d, during which PSP covers 12$^o$ (lower panels f to h). We find that a periodicity between 0.3$^o$ and 1.6$^o$ is strongly present, further confirming our result for $E_5$. As before we overplot in panel \ref{fig: 4_E2}f the IFFT of this peak (shaded in light blue in panel \ref{fig: 4_E2}f), that follows closely the solid angle mid-scale fluctuations as in Figure \ref{fig: 3_space}b.
\section{Discussion and Conclusion} \label{sec:5_disc}

The full time window wavelet analyses of the encounters 1, 2, 4 and 5 are available in the appendix, both over time and space. They are consistent with the above findings, the most striking periods being the ones detailed in sections \ref{sec:3_enc5} and \ref{sec:4_enc2}. We now compare the observed scales to those expected from proposed potential formation process.

\subsection{Turbulent generation of switchbacks}
It has been proposed that switchbacks may form as the solar wind evolves, being produced by turbulence or velocity shears (\cite{Squire_2020}, \cite{Ruffolo_2020}). This is supported by the studies of \cite{Mozer_2020} and \cite{MacNeil_2020} who found that the occurrence of switchbacks increases with radial distance from the Sun $r$. While we cannot conclude on switchback occurrence at radial distances greater than 60~$R_s$, our analysis suggests that it is unrelated to heliocentric radial distance near Sun (see sections \ref{sec: 3.2_time} and \ref{sec:4_enc2}). This is also visible in all encounters (see plots in the appendix). This is at odds with \cite{Mozer_2020}'s results which were based on the comparison of two days of data. Based on our analysis and its extension to four encounters, we rather propose that the occurrence of switchbacks depends on the solar wind properties and origin.

In a wider perspective, the solar wind turbulent cascade in magnetic fluctuations is expected to behave as a rather smooth power law of exponent between $-3/2$ and  $-5/3$ in the inertial range, and as a $-1$ power law at lower frequencies (\cite{Matteini_2019}, \cite{Chen_2020}, \cite{2020_DDW}). In our temporal analysis (Figure \ref{fig: 2_time}), the significant periods we detect start from 2~h, and are well below the break frequency of 0.001~Hz (\~17 minutes) found by \cite{2020_DDW}. They more likely correspond to large spatial structures in the injection scales above the inertial range. We thus suggest that the modulation of the signal in large patches of switchbacks (3$^o$ - 9$^o$) and the remarkable intermediate scale modulation (0.3$^o$ - 1.6$^o$) are not part of the turbulent cascade, although they may contain significant energy available for the turbulence cascade. By contrast the omnipresent high-frequency fluctuations which are more homogeneous may be part of the main turbulence cascade that is expected to be more isotropic.

\subsection{Comparison to granulation and supergranulation}

\begin{figure*}
    \centering
    \includegraphics[width=1\textwidth]{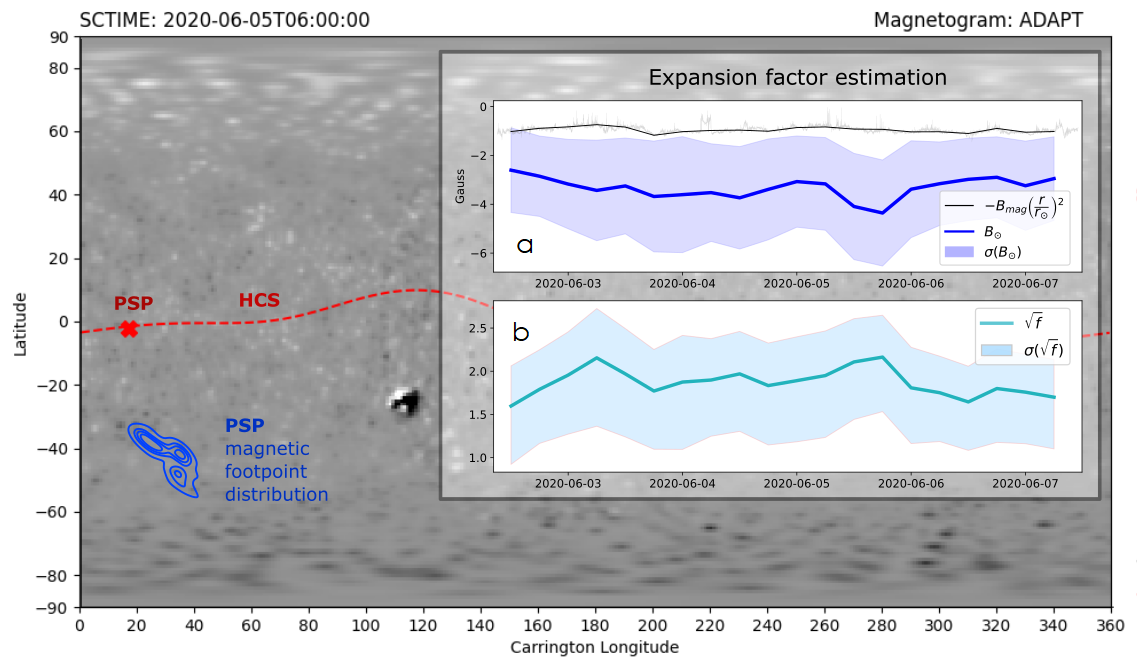}
    \caption{Connectivity analysis performed with the MADAWG connectivity tool$^1$, displaying in blue the distribution of PSP magnetic footpoints over the 5.1-day period analyzed for $E_5$, with an ADAPT magnetogram as a background for context. Panel a) the footpoint magnetic field measured over time (solid blue) with its uncertainty (blue shade), and compared to the one measured by PSP (in black). The square root of the derived expansion factor is then plotted over time in panel b) with its uncertainty (light blue shade).}
    \label{fig:5_connect}
\end{figure*}

The Sun's supergranulation structure and dynamics are not well understood as of today. Using either dopplergrams, tessellation techniques or helioseismology, its typical spatial scale is consistently found around 30~Mega-meters (Mm), with a distribution in size ranging from 20 to 75~Mm (see \cite{2010LRSP....7....2R} and references therein). This range corresponds (by dividing by the Sun's radius) to a typical angular size 1.6$^o$ to 6.2$^o$. Solar granulation is well explained by convective heat transport at the Sun's surface, and presents a typical scale of 1~Mm which yields a 0.08$^o$ angular size
\citep{2009LRSP....6....2N}. The lifetimes of supergranules and granules are respectively around 24h and 10 minutes. Finally, what has sometimes been coined as mesogranulation with an intermediate scale, is now believed to be an artefact of detection techniques (\cite{2009A&A...504.1041M}, \cite{2010A&A...512A...4R}).

In our work, we find significant power in the fluctuation WPS for spatial scales comprised between
[0.3$^o$ - 1.6 $^o$] and [2.6$^o$ - 10.7$^o$]
both in $E_5$ (\ref{sec:3.3_space}) and $E_2$ (\ref{sec:4_enc2}). At a first glance, our values are  larger than those of granulation [0.08$^o$] and supergranulation [1.6$^o$ - 6.2$^o$]. Under the assumption that there is a link between the scales we find and those of granulation and supergranulation, this discrepancy may be explained by the spacecraft connectivity. In our analysis we use the raw projection of the spacecraft position on the Carrington map, hence landing around the equator. However, latitude plays a role when converting distance covered on a flat map to distance covered on a sphere, as highlighted in equation \ref{eq: ds}. To estimate the actual latitude where PSP is connected, we use the connectivity tool developed by the \emph{Solar Orbiter Data Analysis Working Group} (MADAWG) \citep{Rouillard_2020b} and accessible at this website\footnote{\href{http://connect-tool.irap.omp.eu/}{http://connect-tool.irap.omp.eu/}}, tracing field lines to the Sun with PFSS (potential field source surface) modeling. We thus determine that throughout the interval we study for $E_5$, the spacecraft is most probably connected to a latitude between -33$^o$ to -57$^o$ as indicated in Figure \ref{fig:5_connect}. When we run our analysis with $s$ computed at a 40$^o$ latitude, our characteristic scales for $E_5$ become [0.2 - 1.3]$^o$ and [2.0-8.3]$^o$

In addition, it may be argued that the super expansion of the solar wind can lead to an underestimation of expected convection scale sizes at the spacecraft. To assess this we use the connectivity tool cited above with ADAPT magnetograms to determine $B_{\odot}$ the solar surface magnetic field, which is plotted over time in Figure \ref{fig:5_connect}a. We compare it to the value measured by PSP $|B| \left(\dfrac{r}{r_{\odot}}\right)^2 $ (in black in panel \ref{fig:5_connect}a) and derive the expansion factor $f= \dfrac{B_{\odot}}{|B|} \left(\dfrac{r_{\odot}}{r}\right)^2$ (e.g. \cite{2021AA_STANSBY}), which is on average 3.5$\pm$2 over this period. Since $f$ is a ratio between surfaces and considering that we compare characteristic lengths, our detected scales should be divided by a factor $\sqrt{f} = 1.9 \pm 0.6$ (panel \ref{fig:5_connect}b) to be compared to surface processes, yielding [0.12 - 0.7]$^o$ and [1.1 - 4.4]$^o$. 
All of these values are summarized within table \ref{tab:scales} and can be converted to Mm by converting to radians and multiplying by $r_{\odot}$. 
We conclude that the large scales we detect for switchback patches are compatible with supergranulation scales, and that the smaller scales between remain slightly larger than granulation size.

\begin{table}
    \centering
    \begin{tabular}{c|c|c}
    \textbf{Expected scales} & \textbf{granulation} & \textbf{supergranulation} \\
    & ($^o$) & ($^o$) \\
     & 0.08 & [1.6 - 6.2]\\
    \hline
    \hline
    \textbf{Detected scales} & \textbf{medium} & \textbf{large} \\
     & ($^o$) & ($^o$) \\
    0$^o$ lat & [0.3- 1.6] & [2.6 - 10.7]\\
    40$^o$ lat & [0.2- 1.3] & [2.0- 8.3]\\
     40$^o$ lat + expansion & [0.12- 0.7] & [1.1 - 4.4]\\
  
    \end{tabular}
    \caption{Summary of detected scales and expected scales for granulation and supergranulation under various assumptions. All values can be converted to Mm by converting to radians and multiplying by $r_{\odot}$}
    \label{tab:scales}
\end{table}

\subsection{Proposed origin of switchback patches}
\begin{figure}
    \centering
    \includegraphics[width=0.5\textwidth]{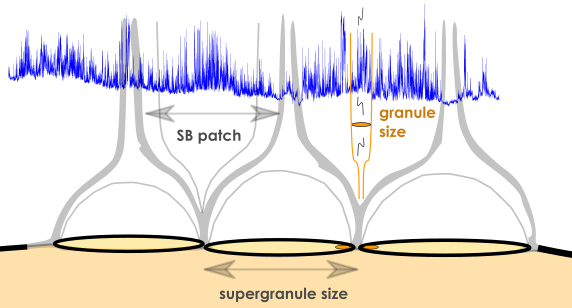}
    \caption{Illustration of swithback modulation by granules and supergranules, with the quantity $B_R(s)$ ($E_5$) overploted in blue for clarity. Grey lines denote magnetic field lines, with thicker ones indicating the separation between closed and open field lines}
    \label{fig:5_schema}
\end{figure}

The modulation of switchback occurrence in patches matching the supergranulation scales and to a lesser extent the granulation scale leads us to believe that at least significant part of the observed switchbacks are produced in the low atmosphere of the Sun. Their occurrence frequency may then indeed be spatially structured by granules and supergranules, as open field lines are rooted at their boundaries. It is interesting to note that \cite{Thieme_1989} found signatures of spatial variations between 2 and 8 degrees in Helios data between 0.3 and 1~AU, though they analyzed variations in density and velocity rather than magnetic field fluctuations. They also underline that the spatial signatures they found were clearer below 0.7 AU, suggesting a solar origin and making a parallel to solar supergranulation. In future work it would be of interest to see if a similar variation in plasma parameters is seen by PSP in association with switchback patches.

We propose an illustration that associates spatial scales of patches and surface structures in Figure \ref{fig:5_schema}. We observe in addition that the background $B_R$ is modulated by the supergranular size, which may be consistent in the overall change in expansion factor within supergranules. The fact that granulation and supergranulation are omnipresent at the Sun's surface while calm solar wind periods devoid of switchbacks are sometimes observed by PSP (\cite{2019Natur.576..237B}, \cite{2020_DDW}, \cite{Malaspina_2020}) is not in contradiction with our result. Indeed, if convection modulates the switchback phenomenon, switchback generation itself might still depend on local surface conditions that remain to be determined, and it is also possible that temporal dependencies arise in their formation  and contribute to the observed trends (although a fortuitous correlation with granular and supergranular scales appears unlikely). The solar wind evolution to PSP might also damp the switchbacks in some places, for yet unknown reasons. The lifetime of granules and supergranules are not relevant in this view, as they would not affect the spatial scales of patches detected by PSP. To conclude, we believe that the fluctuations we observe in switchback occurrence are a superposition of several phenomena: high frequency fluctuations generated at the Sun's surface either through interchange reconnection or by turbulent processes, and larger scales spatial modulations by both supergranulation and granulation, both seen if looking at Figure \ref{fig:5_schema}.

\subsection{Limitations}
In this work, we use the direct projection of the spacecraft orbit on the Carrington surface of the Sun to determine the spatial projection of our data. This relies strongly on the hypothesis that the magnetic footpoints of PSP all follow a similar and linear path, and this is of course not fully adequate as the connection jumps from one source region to another. It is also possible that while the spacecraft skims the edge of a coronal hole, jumps in longitude or latitude occur. To avoid these pitfalls we focused on intervals where, based on in-situ measurements, PFSS modelling, white light analysis and past work, the spacecraft was thought to remain connected to the same source region. Instead of a direct spatial projection we also could have used a ballistic projection of the Parker spiral on the Sun's surface. This technique, however, is poorly suited for spectral analysis, as while the radial velocity of the wind changes, the spiral footpoint can turn around, hence losing the bijection between time and space and folding the signal over itself. In addition, Figure \ref{fig: 1_context} shows that at least for $E_5$ the velocity is not changing much during the interval, making the use of Parker spiral connectivity most likely of little impact.

Another point that is not taken into account with our method is that if the source to which PSP is connected has a limited size, like the small equatorial hole in E1 (\cite{2019Natur.576..237B}, \cite{2020ApJS..246...23B}), the resulting footpoint path could be significantly smaller than the projected orbit. This is actually consistent with the spatial analysis performed over E1 (see appendix) where both the large-scale and mid-scale modulations appear to have lower wavelengths than for $E_5$. To this extent, in future work, it will be interesting to model more precisely the path of the satellite's magnetic footpoint for this type of analysis.

\subsection{Conclusion}
We investigated the phenomena known as magnetic switchbacks observed by PSP, which are interpreted as localised folds in the magnetic field lines. We define switchbacks as a deviation to the Parker spiral and implement an automatic detection on the solid angle between the Parker spiral and the measured magnetic field. We investigate both their temporal and spatial characteristics, using the spacecraft path in curvilinear abscissa $s$, to work in the frame of a spatial projection (expressed in degrees in Carrington coordinates, see Figure \ref{fig: 1_context}). We perform a wavelet analysis on the solid angle fluctuation, focusing on a 5-day period during $E_5$ and on the complete second encounter of PSP with the Sun. We find that :
\begin{itemize}
    \item The detected temporal scales vary over time but do not obviously repeat in a coherent manner throughout the 5-day interval of $E_5$ or the other encounters (see Figure \ref{fig: 2_time} and appendix). Large patches of switchback last from 5 to 18 hours.
    \item By contrast, significant and persistent local spatial scales are detected throughout the 5-day interval studied on E5. They are also found during E2. Large patches of switchbacks present typical spatial extent of 2.6 to 10.7 degrees. The analysis also underlines switchback patches of intermediate scales between 0.3 and 1.6 degrees that appear consistently throughout the encounter (see Figures \ref{fig: 3_space} and \ref{fig: 4_E2}).
    \item Switchback occurrence and spectral properties seem to depend on the source region of the solar wind rather than on the radial distance of PSP. In $E_2$, the power spectrum of the signal was found to be lower in the coronal hole plasma compared to the streamer plasma, even though the dominant scales remained the same (see Figure \ref{fig: 4_E2} and appendix).
\end{itemize}
The wavelengths we detect are outside of the turbulence inertial range and cover lower frequencies, they more likely correspond to large spatial structures in the injection range. When we compare them to the scales of solar granulation and supergranulation, we obtain values that are consistent to both phenomenon. We conclude that supergranulation and granulation may be respectively the source of the large scale modulation of switchbacks called switchback patches, and the reported mid scale modulation. While we can not conclude on the physical process at stake regarding individual switchback formation (magnetic reconnection, turbulence or yet another process), our result nevertheless suggest that switchbacks most probably originate in the low solar atmosphere since their occurrence appears to be modulated by the effects of solar surface motion at the granular and supergranular scales.

\acknowledgments

Work at IRAP was supported by CNRS, CNES and UPS. CF and TD also acknowledge support from CNES. We acknowledge the NASA Parker Solar Probe Mission and particularly the FIELDS team led by S. D. Bale and the SWEAP team led by J. Kasper for use of data. Parker Solar Probe was designed, built, and is now operated by the Johns Hopkins Applied Physics Laboratory as part of NASA’s Living with a Star (LWS) program (contract NNN06AA01C). Support from the LWS management and technical team has played a critical role in the success of the Parker Solar Probe mission. The data used in this study are available at the NASA Space Physics Data Facility (SPDF): https://spdf.gsfc.nasa.gov, and at the PSP science gateway https://sppgway.jhuapl.edu/. We visualized data using the CLWeb software available at http://clweb.irap.omp.eu/, developed by E. Penou. The author NF acknowledges the support of the ISSI team, working in unraveling solar wind microphysics in the inner heliosphere.

\bibliography{SOURCES}{}

\begin{thebibliography}{}
\expandafter\ifx\csname natexlab\endcsname\relax\def\natexlab#1{#1}\fi
\providecommand{\url}[1]{\href{#1}{#1}}
\providecommand{\dodoi}[1]{doi:~\href{http://doi.org/#1}{\nolinkurl{#1}}}
\providecommand{\doeprint}[1]{\href{http://ascl.net/#1}{\nolinkurl{http://ascl.net/#1}}}
\providecommand{\doarXiv}[1]{\href{https://arxiv.org/abs/#1}{\nolinkurl{https://arxiv.org/abs/#1}}}

\bibitem[{{Auch{\`e}re} {et~al.}(2016){Auch{\`e}re}, {Froment}, {Bocchialini},
  {Buchlin}, \& {Solomon}}]{2016ApJ...825..110A}
{Auch{\`e}re}, F., {Froment}, C., {Bocchialini}, K., {Buchlin}, E., \&
  {Solomon}, J. 2016, \apj, 825, 110, \dodoi{10.3847/0004-637X/825/2/110}

\bibitem[{{Badman} {et~al.}(2020){Badman}, {Bale}, {Mart{\'\i}nez Oliveros},
  {Panasenco}, {Velli}, {Stansby}, {Buitrago-Casas}, {R{\'e}ville}, {Bonnell},
  {Case}, {Dudok de Wit}, {Goetz}, {Harvey}, {Kasper}, {Korreck}, {Larson},
  {Livi}, {MacDowall}, {Malaspina}, {Pulupa}, {Stevens}, \&
  {Whittlesey}}]{2020ApJS..246...23B}
{Badman}, S.~T., {Bale}, S.~D., {Mart{\'\i}nez Oliveros}, J.~C., {et~al.} 2020,
  \apjs, 246, 23, \dodoi{10.3847/1538-4365/ab4da7}

\bibitem[{{Bale} {et~al.}(2016){Bale}, {Goetz}, {Harvey}, {Turin}, {Bonnell},
  {Dudok de Wit}, {Ergun}, {MacDowall}, {Pulupa}, {Andre}, {Bolton},
  {Bougeret}, {Bowen}, {Burgess}, {Cattell}, {Chandran}, {Chaston}, {Chen},
  {Choi}, {Connerney}, {Cranmer}, {Diaz-Aguado}, {Donakowski}, {Drake},
  {Farrell}, {Fergeau}, {Fermin}, {Fischer}, {Fox}, {Glaser}, {Goldstein},
  {Gordon}, {Hanson}, {Harris}, {Hayes}, {Hinze}, {Hollweg}, {Horbury},
  {Howard}, {Hoxie}, {Jannet}, {Karlsson}, {Kasper}, {Kellogg}, {Kien},
  {Klimchuk}, {Krasnoselskikh}, {Krucker}, {Lynch}, {Maksimovic}, {Malaspina},
  {Marker}, {Martin}, {Martinez-Oliveros}, {McCauley}, {McComas}, {McDonald},
  {Meyer-Vernet}, {Moncuquet}, {Monson}, {Mozer}, {Murphy}, {Odom},
  {Oliverson}, {Olson}, {Parker}, {Pankow}, {Phan}, {Quataert}, {Quinn},
  {Ruplin}, {Salem}, {Seitz}, {Sheppard}, {Siy}, {Stevens}, {Summers}, {Szabo},
  {Timofeeva}, {Vaivads}, {Velli}, {Yehle}, {Werthimer}, \&
  {Wygant}}]{2016SSRv..204...49B}
{Bale}, S.~D., {Goetz}, K., {Harvey}, P.~R., {et~al.} 2016, \ssr, 204, 49,
  \dodoi{10.1007/s11214-016-0244-5}

\bibitem[{{Bale} {et~al.}(2019){Bale}, {Badman}, {Bonnell}, {Bowen}, {Burgess},
  {Case}, {Cattell}, {Chandran}, {Chaston}, {Chen}, {Drake}, {de Wit},
  {Eastwood}, {Ergun}, {Farrell}, {Fong}, {Goetz}, {Goldstein}, {Goodrich},
  {Harvey}, {Horbury}, {Howes}, {Kasper}, {Kellogg}, {Klimchuk}, {Korreck},
  {Krasnoselskikh}, {Krucker}, {Laker}, {Larson}, {MacDowall}, {Maksimovic},
  {Malaspina}, {Martinez-Oliveros}, {McComas}, {Meyer-Vernet}, {Moncuquet},
  {Mozer}, {Phan}, {Pulupa}, {Raouafi}, {Salem}, {Stansby}, {Stevens}, {Szabo},
  {Velli}, {Woolley}, \& {Wygant}}]{2019Natur.576..237B}
{Bale}, S.~D., {Badman}, S.~T., {Bonnell}, J.~W., {et~al.} 2019, \nat, 576,
  237, \dodoi{10.1038/s41586-019-1818-7}

\bibitem[{{Balogh} {et~al.}(1999){Balogh}, {Forsyth}, {Lucek}, {Horbury}, \&
  {Smith}}]{1999GeoRL..26..631B}
{Balogh}, A., {Forsyth}, R.~J., {Lucek}, E.~A., {Horbury}, T.~S., \& {Smith},
  E.~J. 1999, \grl, 26, 631, \dodoi{10.1029/1999GL900061}

\bibitem[{{Case} {et~al.}(2020){Case}, {Kasper}, {Stevens}, {Korreck},
  {Paulson}, {Daigneau}, {Caldwell}, {Freeman}, {Henry}, {Klingensmith},
  {Bookbinder}, {Robinson}, {Berg}, {Tiu}, {Wright}, {Reinhart}, {Curtis},
  {Ludlam}, {Larson}, {Whittlesey}, {Livi}, {Klein}, \&
  {Martinovi{\'c}}}]{2020ApJS..246...43C}
{Case}, A.~W., {Kasper}, J.~C., {Stevens}, M.~L., {et~al.} 2020, \apjs, 246,
  43, \dodoi{10.3847/1538-4365/ab5a7b}

\bibitem[{{Chen} {et~al.}(2020){Chen}, {Bale}, {Bonnell}, {Borovikov}, {Bowen},
  {Burgess}, {Case}, {Chandran}, {de Wit}, {Goetz}, {Harvey}, {Kasper},
  {Klein}, {Korreck}, {Larson}, {Livi}, {MacDowall}, {Malaspina}, {Mallet},
  {McManus}, {Moncuquet}, {Pulupa}, {Stevens}, \& {Whittlesey}}]{Chen_2020}
{Chen}, C.~H.~K., {Bale}, S.~D., {Bonnell}, J.~W., {et~al.} 2020, \apjs, 246,
  53, \dodoi{10.3847/1538-4365/ab60a3}

\bibitem[{{Cranmer} {et~al.}(2007){Cranmer}, {van Ballegooijen}, \&
  {Edgar}}]{2007ApJS..171..520C}
{Cranmer}, S.~R., {van Ballegooijen}, A.~A., \& {Edgar}, R.~J. 2007, \apjs,
  171, 520, \dodoi{10.1086/518001}

\bibitem[{{Dierckx}(1993)}]{1993csfw.book.....D}
{Dierckx}, P. 1993, {Curve and surface fitting with splines}

\bibitem[{{Drake} {et~al.}(2021){Drake}, {Agapitov}, {Swisdak}, {Badman},
  {Bale}, {Horbury}, {Kasper}, {MacDowall}, {Mozer}, {Phan}, {Pulupa}, {Szabo},
  \& {Velli}}]{2021AA_DRAKE}
{Drake}, J.~F., {Agapitov}, O., {Swisdak}, M., {et~al.} 2021, \aap, 650, A2,
  \dodoi{10.1051/0004-6361/202039432}

\bibitem[{{Dudok de Wit} {et~al.}(2020){Dudok de Wit}, {Krasnoselskikh},
  {Bale}, {Bonnell}, {Bowen}, {Chen}, {Froment}, {Goetz}, {Harvey},
  {Jagarlamudi}, {Larosa}, {MacDowall}, {Malaspina}, {Matthaeus}, {Pulupa},
  {Velli}, \& {Whittlesey}}]{2020_DDW}
{Dudok de Wit}, T., {Krasnoselskikh}, V.~V., {Bale}, S.~D., {et~al.} 2020,
  \apjs, 246, 39, \dodoi{10.3847/1538-4365/ab5853}

\bibitem[{{Fargette} {et~al.}(2021){Fargette}, {Lavraud}, {Rouillard},
  {Eastwood}, {Bale}, {Phan}, {{\O}ieroset}, {Halekas}, {Kasper}, {Berthomier},
  {Case}, {Korreck}, {Larson}, {Louarn}, {Malaspina}, {Pulupa}, {Stevens},
  {Whittlesey}, {MacDowall}, {Goetz}, {Harvey}, {Dudok de Wit}, \&
  {Bonnell}}]{2021AA_FARGETTE}
{Fargette}, N., {Lavraud}, B., {Rouillard}, A., {et~al.} 2021, \aap, 650, A11,
  \dodoi{10.1051/0004-6361/202039191}

\bibitem[{{Fisk}(2005)}]{2005ApJ...626..563F}
{Fisk}, L.~A. 2005, \apj, 626, 563, \dodoi{10.1086/429957}

\bibitem[{{Fisk} \& {Kasper}(2020)}]{2020ApJ...894L...4F}
{Fisk}, L.~A., \& {Kasper}, J.~C. 2020, \apjl, 894, L4,
  \dodoi{10.3847/2041-8213/ab8acd}

\bibitem[{{Fisk} \& {Schwadron}(2001)}]{2001ApJ...560..425F}
{Fisk}, L.~A., \& {Schwadron}, N.~A. 2001, \apj, 560, 425,
  \dodoi{10.1086/322503}

\bibitem[{{Gosling} {et~al.}(2006){Gosling}, {McComas}, {Skoug}, \&
  {Smith}}]{Gosling_2006}
{Gosling}, J.~T., {McComas}, D.~J., {Skoug}, R.~M., \& {Smith}, C.~W. 2006,
  \grl, 33, L17102, \dodoi{10.1029/2006GL027188}

\bibitem[{{Gosling} {et~al.}(2011){Gosling}, {Tian}, \&
  {Phan}}]{2011ApJ...737L..35G}
{Gosling}, J.~T., {Tian}, H., \& {Phan}, T.~D. 2011, \apjl, 737, L35,
  \dodoi{10.1088/2041-8205/737/2/L35}

\bibitem[{{Griton} {et~al.}(2021){Griton}, {Rouillard}, {Poirier}, {Issautier},
  {Moncuquet}, \& {Pinto}}]{Griton_2021}
{Griton}, L., {Rouillard}, A.~P., {Poirier}, N., {et~al.} 2021, \apj, 910, 63,
  \dodoi{10.3847/1538-4357/abe309}

\bibitem[{{Horbury} {et~al.}(2018){Horbury}, {Matteini}, \&
  {Stansby}}]{2018MNRAS.478.1980H}
{Horbury}, T.~S., {Matteini}, L., \& {Stansby}, D. 2018, \mnras, 478, 1980,
  \dodoi{10.1093/mnras/sty953}

\bibitem[{{Horbury} {et~al.}(2020){Horbury}, {Woolley}, {Laker}, {Matteini},
  {Eastwood}, {Bale}, {Velli}, {Chandran}, {Phan}, {Raouafi}, {Goetz},
  {Harvey}, {Pulupa}, {Klein}, {Dudok de Wit}, {Kasper}, {Korreck}, {Case},
  {Stevens}, {Whittlesey}, {Larson}, {MacDowall}, {Malaspina}, \&
  {Livi}}]{2020ApJS..246...45H}
{Horbury}, T.~S., {Woolley}, T., {Laker}, R., {et~al.} 2020, \apjs, 246, 45,
  \dodoi{10.3847/1538-4365/ab5b15}

\bibitem[{{Kasper} {et~al.}(2016){Kasper}, {Abiad}, {Austin}, {Balat-Pichelin},
  {Bale}, {Belcher}, {Berg}, {Bergner}, {Berthomier}, {Bookbinder}, {Brodu},
  {Caldwell}, {Case}, {Chand ran}, {Cheimets}, {Cirtain}, {Cranmer}, {Curtis},
  {Daigneau}, {Dalton}, {Dasgupta}, {DeTomaso}, {Diaz-Aguado}, {Djordjevic},
  {Donaskowski}, {Effinger}, {Florinski}, {Fox}, {Freeman}, {Gallagher},
  {Gary}, {Gauron}, {Gates}, {Goldstein}, {Golub}, {Gordon}, {Gurnee}, {Guth},
  {Halekas}, {Hatch}, {Heerikuisen}, {Ho}, {Hu}, {Johnson}, {Jordan},
  {Korreck}, {Larson}, {Lazarus}, {Li}, {Livi}, {Ludlam}, {Maksimovic},
  {McFadden}, {Marchant}, {Maruca}, {McComas}, {Messina}, {Mercer}, {Park},
  {Peddie}, {Pogorelov}, {Reinhart}, {Richardson}, {Robinson}, {Rosen},
  {Skoug}, {Slagle}, {Steinberg}, {Stevens}, {Szabo}, {Taylor}, {Tiu}, {Turin},
  {Velli}, {Webb}, {Whittlesey}, {Wright}, {Wu}, \&
  {Zank}}]{2016SSRv..204..131K}
{Kasper}, J.~C., {Abiad}, R., {Austin}, G., {et~al.} 2016, \ssr, 204, 131,
  \dodoi{10.1007/s11214-015-0206-3}

\bibitem[{{Kasper} {et~al.}(2019){Kasper}, {Bale}, {Belcher}, {Berthomier},
  {Case}, {Chandran}, {Curtis}, {Gallagher}, {Gary}, {Golub}, {Halekas}, {Ho},
  {Horbury}, {Hu}, {Huang}, {Klein}, {Korreck}, {Larson}, {Livi}, {Maruca},
  {Lavraud}, {Louarn}, {Maksimovic}, {Martinovic}, {McGinnis}, {Pogorelov},
  {Richardson}, {Skoug}, {Steinberg}, {Stevens}, {Szabo}, {Velli},
  {Whittlesey}, {Wright}, {Zank}, {MacDowall}, {McComas}, {McNutt}, {Pulupa},
  {Raouafi}, \& {Schwadron}}]{2019Natur.576..228K}
{Kasper}, J.~C., {Bale}, S.~D., {Belcher}, J.~W., {et~al.} 2019, \nat, 576,
  228, \dodoi{10.1038/s41586-019-1813-z}

\bibitem[{{Korreck} {et~al.}(2020){Korreck}, {Szabo}, {Nieves Chinchilla},
  {Lavraud}, {Luhmann}, {Niembro}, {Higginson}, {Alzate}, {Wallace}, {Paulson},
  {Rouillard}, {Kouloumvakos}, {Poirier}, {Kasper}, {Case}, {Stevens}, {Bale},
  {Pulupa}, {Whittlesey}, {Livi}, {Goetz}, {Larson}, {Malaspina}, {Morgan},
  {Narock}, {Schwadron}, {Bonnell}, {Harvey}, \&
  {Wygant}}]{2020ApJS..246...69K}
{Korreck}, K.~E., {Szabo}, A., {Nieves Chinchilla}, T., {et~al.} 2020, \apjs,
  246, 69, \dodoi{10.3847/1538-4365/ab6ff9}

\bibitem[{{Laker} {et~al.}(2021){Laker}, {Horbury}, {Bale}, {Matteini},
  {Woolley}, {Woodham}, {Badman}, {Pulupa}, {Kasper}, {Stevens}, {Case}, \&
  {Korreck}}]{2021AA_LAKER}
{Laker}, R., {Horbury}, T.~S., {Bale}, S.~D., {et~al.} 2021, \aap, 650, A1,
  \dodoi{10.1051/0004-6361/202039354}

\bibitem[{{Lavraud} {et~al.}(2020){Lavraud}, {Fargette}, {R{\'e}ville},
  {Szabo}, {Huang}, {Rouillard}, {Viall}, {Phan}, {Kasper}, {Bale},
  {Berthomier}, {Bonnell}, {Case}, {Dudok de Wit}, {Eastwood}, {G{\'e}not},
  {Goetz}, {Griton}, {Halekas}, {Harvey}, {Kieokaew}, {Klein}, {Korreck},
  {Kouloumvakos}, {Larson}, {Lavarra}, {Livi}, {Louarn}, {MacDowall},
  {Maksimovic}, {Malaspina}, {Nieves-Chinchilla}, {Pinto}, {Poirier}, {Pulupa},
  {Raouafi}, {Stevens}, {Toledo-Redondo}, \&
  {Whittlesey}}]{2020ApJ...894L..19L}
{Lavraud}, B., {Fargette}, N., {R{\'e}ville}, V., {et~al.} 2020, \apjl, 894,
  L19, \dodoi{10.3847/2041-8213/ab8d2d}

\bibitem[{{Macneil} {et~al.}(2020){Macneil}, {Owens}, {Wicks}, {Lockwood},
  {Bentley}, \& {Lang}}]{MacNeil_2020}
{Macneil}, A.~R., {Owens}, M.~J., {Wicks}, R.~T., {et~al.} 2020, \mnras, 494,
  3642, \dodoi{10.1093/mnras/staa951}

\bibitem[{{Malaspina} {et~al.}(2020){Malaspina}, {Halekas},
  {Ber{\v{c}}i{\v{c}}}, {Larson}, {Whittlesey}, {Bale}, {Bonnell}, {Dudok de
  Wit}, {Ergun}, {Howes}, {Goetz}, {Goodrich}, {Harvey}, {MacDowall}, {Pulupa},
  {Case}, {Kasper}, {Korreck}, {Livi}, \& {Stevens}}]{Malaspina_2020}
{Malaspina}, D.~M., {Halekas}, J., {Ber{\v{c}}i{\v{c}}}, L., {et~al.} 2020,
  \apjs, 246, 21, \dodoi{10.3847/1538-4365/ab4c3b}

\bibitem[{{Matloch} {et~al.}(2009){Matloch}, {Cameron}, {Schmitt}, \&
  {Sch{\"u}ssler}}]{2009A&A...504.1041M}
{Matloch}, L., {Cameron}, R., {Schmitt}, D., \& {Sch{\"u}ssler}, M. 2009, \aap,
  504, 1041, \dodoi{10.1051/0004-6361/200811200}

\bibitem[{{Matteini} {et~al.}(2014){Matteini}, {Horbury}, {Neugebauer}, \&
  {Goldstein}}]{Matteini_2014}
{Matteini}, L., {Horbury}, T.~S., {Neugebauer}, M., \& {Goldstein}, B.~E. 2014,
  \grl, 41, 259, \dodoi{10.1002/2013GL058482}

\bibitem[{{Matteini} {et~al.}(2019){Matteini}, {Stansby}, {Horbury}, \&
  {Chen}}]{Matteini_2019}
{Matteini}, L., {Stansby}, D., {Horbury}, T.~S., \& {Chen}, C.~H.~K. 2019,
  Nuovo Cimento C Geophysics Space Physics C, 42, 16,
  \dodoi{10.1393/ncc/i2019-19016-y}

\bibitem[{{Meyer-Vernet}(2007)}]{MEYER_2007_book}
{Meyer-Vernet}, N. 2007, {Basics of the Solar Wind}

\bibitem[{{Mozer} {et~al.}(2020){Mozer}, {Agapitov}, {Bale}, {Bonnell}, {Case},
  {Chaston}, {Curtis}, {Dudok de Wit}, {Goetz}, {Goodrich}, {Harvey}, {Kasper},
  {Korreck}, {Krasnoselskikh}, {Larson}, {Livi}, {MacDowall}, {Malaspina},
  {Pulupa}, {Stevens}, {Whittlesey}, \& {Wygant}}]{Mozer_2020}
{Mozer}, F.~S., {Agapitov}, O.~V., {Bale}, S.~D., {et~al.} 2020, \apjs, 246,
  68, \dodoi{10.3847/1538-4365/ab7196}

\bibitem[{{Neugebauer} \& {Goldstein}(2013)}]{Neugebauer_2013}
{Neugebauer}, M., \& {Goldstein}, B.~E. 2013, in American Institute of Physics
  Conference Series, Vol. 1539, Solar Wind 13, ed. G.~P. {Zank}, J.~{Borovsky},
  R.~{Bruno}, J.~{Cirtain}, S.~{Cranmer}, H.~{Elliott}, J.~{Giacalone},
  W.~{Gonzalez}, G.~{Li}, E.~{Marsch}, E.~{Moebius}, N.~{Pogorelov},
  J.~{Spann}, \& O.~{Verkhoglyadova}, 46--49, \dodoi{10.1063/1.4810986}

\bibitem[{{Nieves-Chinchilla} {et~al.}(2020){Nieves-Chinchilla}, {Szabo},
  {Korreck}, {Alzate}, {Balmaceda}, {Lavraud}, {Paulson}, {Narock}, {Wallace},
  {Jian}, {Luhmann}, {Morgan}, {Higginson}, {Arge}, {Bale}, {Case}, {Dudfok de
  Wit}, {Giacalone}, {Goetz}, {Harvey}, {Jones-Melosky}, {Kasper}, {Larson},
  {Livi}, {McComas}, {MacDowall}, {Malaspina}, {Pulupa}, {Raouafi},
  {Schwadron}, {Stevens}, \& {Whittlesey}}]{2020ApJS..246...63N}
{Nieves-Chinchilla}, T., {Szabo}, A., {Korreck}, K.~E., {et~al.} 2020, \apjs,
  246, 63, \dodoi{10.3847/1538-4365/ab61f5}

\bibitem[{{Nordlund} {et~al.}(2009){Nordlund}, {Stein}, \&
  {Asplund}}]{2009LRSP....6....2N}
{Nordlund}, {\r{A}}., {Stein}, R.~F., \& {Asplund}, M. 2009, Living Reviews in
  Solar Physics, 6, 2, \dodoi{10.12942/lrsp-2009-2}

\bibitem[{{Phan} {et~al.}(2020){Phan}, {Bale}, {Eastwood}, {Lavraud}, {Drake},
  {Oieroset}, {Shay}, {Pulupa}, {Stevens}, {MacDowall}, {Case}, {Larson},
  {Kasper}, {Whittlesey}, {Szabo}, {Korreck}, {Bonnell}, {de Wit}, {Goetz},
  {Harvey}, {Horbury}, {Livi}, {Malaspina}, {Paulson}, {Raouafi}, \&
  {Velli}}]{2020ApJS..246...34P}
{Phan}, T.~D., {Bale}, S.~D., {Eastwood}, J.~P., {et~al.} 2020, \apjs, 246, 34,
  \dodoi{10.3847/1538-4365/ab55ee}

\bibitem[{{R{\'e}ville} {et~al.}(2020){R{\'e}ville}, {Velli}, {Panasenco},
  {Tenerani}, {Shi}, {Badman}, {Bale}, {Kasper}, {Stevens}, {Korreck},
  {Bonnell}, {Case}, {de Wit}, {Goetz}, {Harvey}, {Larson}, {Livi},
  {Malaspina}, {MacDowall}, {Pulupa}, \& {Whittlesey}}]{2020ApJS..246...24R}
{R{\'e}ville}, V., {Velli}, M., {Panasenco}, O., {et~al.} 2020, \apjs, 246, 24,
  \dodoi{10.3847/1538-4365/ab4fef}

\bibitem[{{Rieutord} \& {Rincon}(2010)}]{2010LRSP....7....2R}
{Rieutord}, M., \& {Rincon}, F. 2010, Living Reviews in Solar Physics, 7, 2,
  \dodoi{10.12942/lrsp-2010-2}

\bibitem[{{Rieutord} {et~al.}(2010){Rieutord}, {Roudier}, {Rincon}, {Malherbe},
  {Meunier}, {Berger}, \& {Frank}}]{2010A&A...512A...4R}
{Rieutord}, M., {Roudier}, T., {Rincon}, F., {et~al.} 2010, \aap, 512, A4,
  \dodoi{10.1051/0004-6361/200913303}

\bibitem[{{Roudier} {et~al.}(2009){Roudier}, {Rieutord}, {Brito}, {Rincon},
  {Malherbe}, {Meunier}, {Berger}, \& {Frank}}]{2009A&A...495..945R}
{Roudier}, T., {Rieutord}, M., {Brito}, D., {et~al.} 2009, \aap, 495, 945,
  \dodoi{10.1051/0004-6361:200811101}

\bibitem[{{Rouillard} {et~al.}(2020{\natexlab{a}}){Rouillard}, {Kouloumvakos},
  {Vourlidas}, {Kasper}, {Bale}, {Raouafi}, {Lavraud}, {Howard}, {Stenborg},
  {Stevens}, {Poirier}, {Davies}, {Hess}, {Higginson}, {Lavarra}, {Viall},
  {Korreck}, {Pinto}, {Griton}, {R{\'e}ville}, {Louarn}, {Wu}, {Dalmasse},
  {G{\'e}not}, {Case}, {Whittlesey}, {Larson}, {Halekas}, {Livi}, {Goetz},
  {Harvey}, {MacDowall}, {Malaspina}, {Pulupa}, {Bonnell}, {de Witt}, \&
  {Penou}}]{2020ApJS..246...37R}
{Rouillard}, A.~P., {Kouloumvakos}, A., {Vourlidas}, A., {et~al.}
  2020{\natexlab{a}}, \apjs, 246, 37, \dodoi{10.3847/1538-4365/ab579a}

\bibitem[{{Rouillard} {et~al.}(2020{\natexlab{b}}){Rouillard}, {Pinto},
  {Vourlidas}, {De Groof}, {Thompson}, {Bemporad}, {Dolei}, {Indurain},
  {Buchlin}, {Sasso}, {Spadaro}, {Dalmasse}, {Hirzberger}, {Zouganelis},
  {Strugarek}, {Brun}, {Alexandre}, {Berghmans}, {Raouafi}, {Wiegelmann},
  {Pagano}, {Arge}, {Nieves-Chinchilla}, {Lavarra}, {Poirier}, {Amari}, {Aran},
  {Andretta}, {Antonucci}, {Anastasiadis}, {Auch{\`e}re}, {Bellot Rubio},
  {Nicula}, {Bonnin}, {Bouchemit}, {Budnik}, {Caminade}, {Cecconi}, {Carlyle},
  {Cernuda}, {Davila}, {Etesi}, {Espinosa Lara}, {Fedorov}, {Fineschi},
  {Fludra}, {G{\'e}not}, {Georgoulis}, {Gilbert}, {Giunta}, {Gomez-Herrero},
  {Guest}, {Haberreiter}, {Hassler}, {Henney}, {Howard}, {Horbury}, {Janvier},
  {Jones}, {Kozarev}, {Kraaikamp}, {Kouloumvakos}, {Krucker}, {Lagg}, {Linker},
  {Lavraud}, {Louarn}, {Maksimovic}, {Maloney}, {Mann}, {Masson}, {M{\"u}ller},
  {{\"O}nel}, {Osuna}, {Orozco Suarez}, {Owen}, {Papaioannou},
  {P{\'e}rez-Su{\'a}rez}, {Rodriguez-Pacheco}, {Parenti}, {Pariat}, {Peter},
  {Plunkett}, {Pomoell}, {Raines}, {Riethm{\"u}ller}, {Rich}, {Rodriguez},
  {Romoli}, {Sanchez}, {Solanki}, {St Cyr}, {Straus}, {Susino}, {Teriaca}, {del
  Toro Iniesta}, {Ventura}, {Verbeeck}, {Vilmer}, {Warmuth}, {Walsh}, {Watson},
  {Williams}, {Wu}, \& {Zhukov}}]{Rouillard_2020b}
{Rouillard}, A.~P., {Pinto}, R.~F., {Vourlidas}, A., {et~al.}
  2020{\natexlab{b}}, \aap, 642, A2, \dodoi{10.1051/0004-6361/201935305}

\bibitem[{{Ruffolo} {et~al.}(2020){Ruffolo}, {Matthaeus}, {Chhiber}, {Usmanov},
  {Yang}, {Bandyopadhyay}, {Parashar}, {Goldstein}, {DeForest}, {Wan},
  {Chasapis}, {Maruca}, {Velli}, \& {Kasper}}]{Ruffolo_2020}
{Ruffolo}, D., {Matthaeus}, W.~H., {Chhiber}, R., {et~al.} 2020, \apj, 902, 94,
  \dodoi{10.3847/1538-4357/abb594}

\bibitem[{{Schwadron} \& {McComas}(2021)}]{Schwadron_2021}
{Schwadron}, N.~A., \& {McComas}, D.~J. 2021, \apj, 909, 95,
  \dodoi{10.3847/1538-4357/abd4e6}

\bibitem[{{Squire} {et~al.}(2020){Squire}, {Chandran}, \&
  {Meyrand}}]{Squire_2020}
{Squire}, J., {Chandran}, B.~D.~G., \& {Meyrand}, R. 2020, \apjl, 891, L2,
  \dodoi{10.3847/2041-8213/ab74e1}

\bibitem[{{Stansby} {et~al.}(2021){Stansby}, {Ber{\v{c}}i{\v{c}}}, {Matteini},
  {Owen}, {French}, {Baker}, \& {Badman}}]{2021AA_STANSBY}
{Stansby}, D., {Ber{\v{c}}i{\v{c}}}, L., {Matteini}, L., {et~al.} 2021, \aap,
  650, L2, \dodoi{10.1051/0004-6361/202039789}

\bibitem[{{Szabo} {et~al.}(2020){Szabo}, {Larson}, {Whittlesey}, {Stevens},
  {Lavraud}, {Phan}, {Wallace}, {Jones-Mecholsky}, {Arge}, {Badman},
  {Odstrcil}, {Pogorelov}, {Kim}, {Riley}, {Henney}, {Bale}, {Bonnell}, {Case},
  {Dudok de Wit}, {Goetz}, {Harvey}, {Kasper}, {Korreck}, {Koval}, {Livi},
  {MacDowall}, {Malaspina}, \& {Pulupa}}]{2020ApJS..246...47S}
{Szabo}, A., {Larson}, D., {Whittlesey}, P., {et~al.} 2020, \apjs, 246, 47,
  \dodoi{10.3847/1538-4365/ab5dac}

\bibitem[{{Tenerani} {et~al.}(2020){Tenerani}, {Velli}, {Matteini},
  {R{\'e}ville}, {Shi}, {Bale}, {Kasper}, {Bonnell}, {Case}, {de Wit}, {Goetz},
  {Harvey}, {Klein}, {Korreck}, {Larson}, {Livi}, {MacDowall}, {Malaspina},
  {Pulupa}, {Stevens}, \& {Whittlesey}}]{2020ApJS..246...32T}
{Tenerani}, A., {Velli}, M., {Matteini}, L., {et~al.} 2020, \apjs, 246, 32,
  \dodoi{10.3847/1538-4365/ab53e1}

\bibitem[{{Thieme} {et~al.}(1989){Thieme}, {Schwenn}, \&
  {Marsch}}]{Thieme_1989}
{Thieme}, K.~M., {Schwenn}, R., \& {Marsch}, E. 1989, Advances in Space
  Research, 9, 127, \dodoi{10.1016/0273-1177(89)90105-1}

\bibitem[{{Torrence} \& {Compo}(1998)}]{1998BAMS...79...61T}
{Torrence}, C., \& {Compo}, G.~P. 1998, Bulletin of the American Meteorological
  Society, 79, 61, \dodoi{10.1175/1520-0477(1998)079<0061:APGTWA>2.0.CO;2}

\bibitem[{{van Ballegooijen} \& {Asgari-Targhi}(2016)}]{2016ApJ...821..106V}
{van Ballegooijen}, A.~A., \& {Asgari-Targhi}, M. 2016, \apj, 821, 106,
  \dodoi{10.3847/0004-637X/821/2/106}

\bibitem[{{Velli} {et~al.}(1989){Velli}, {Grappin}, \& {Mangeney}}]{Velli_1989}
{Velli}, M., {Grappin}, R., \& {Mangeney}, A. 1989, \prl, 63, 1807,
  \dodoi{10.1103/PhysRevLett.63.1807}

\bibitem[{{Whittlesey} {et~al.}(2020){Whittlesey}, {Larson}, {Kasper},
  {Halekas}, {Abatcha}, {Abiad}, {Berthomier}, {Case}, {Chen}, {Curtis},
  {Dalton}, {Klein}, {Korreck}, {Livi}, {Ludlam}, {Marckwordt}, {Rahmati},
  {Robinson}, {Slagle}, {Stevens}, {Tiu}, \& {Verniero}}]{2020ApJS..246...74W}
{Whittlesey}, P.~L., {Larson}, D.~E., {Kasper}, J.~C., {et~al.} 2020, \apjs,
  246, 74, \dodoi{10.3847/1538-4365/ab7370}

\bibitem[{{Woolley} {et~al.}(2020){Woolley}, {Matteini}, {Horbury}, {Bale},
  {Woodham}, {Laker}, {Alterman}, {Bonnell}, {Case}, {Kasper}, {Klein},
  {Martinovi{\'c}}, \& {Stevens}}]{Woolley_2020}
{Woolley}, T., {Matteini}, L., {Horbury}, T.~S., {et~al.} 2020, \mnras, 498,
  5524, \dodoi{10.1093/mnras/staa2770}

\bibitem[{{Yamauchi} {et~al.}(2004){Yamauchi}, {Suess}, {Steinberg}, \&
  {Sakurai}}]{yamauchi_2004}
{Yamauchi}, Y., {Suess}, S.~T., {Steinberg}, J.~T., \& {Sakurai}, T. 2004,
  Journal of Geophysical Research (Space Physics), 109, A03104,
  \dodoi{10.1029/2003JA010274}

\end{thebibliography}
\bibliographystyle{aasjournal}

\appendix
We display in Table \ref{tab:intervals} the list of intervals that were discarded in our study, as they were identified as either heliospheric current sheet (HCS) crossings, heliospheric plasma sheets (HPS) crossings, Magnetic Increases with Central Current Sheet (MICCS) structures, coronal mass ejections (CME), or periods of strahl drop out where magnetic field lines are most likely disconnected from the Sun . All of these intervals are identified visually while scanning through the data.

\begin{table*}[h]
    \centering
    \begin{tabular}{c|c|c|c}
    Encounter & Start & End & Type\\
    \hline
    \hline
    1 & 2018-10-31 04:00 &  2018-10-31 12:20 & CME\\
    1 & 2018-11-01 23:00 &  2018-11-01 23:15 & MICCS\\
    1 & 2018-11-02 12:35 &  2018-11-02 12:50 & MICCS\\
    1 & 2018-11-08 23:20 &  2018-11-08 23:45 & MICCS\\
    1 & 2018-11-11 17:00 &  2018-11-12 12:00 & CME\\
    \hline
    4 & 2020-01-28 01:00 &  2020-01-28 01:10 & MICCS\\
    4 & 2020-01-30 13:15 &  2020-01-30 17:10 & Partial HPS crossing with strahl drop out\\
    4 & 2020-01-31 19:50 &  2020-02-01 00:05 & Partial HPS crossing with strahl drop out\\
    4 & 2020-02-01 03:55 &  2020-02-01 04:15 & HCS crossing\\
    \hline
    5 & 2020-05-31 12:21 &  2020-06-01 03:40 & Flux rope, MICCS and strahl drop out\\
    5 & 2020-06-01 10:00 &  2020-06-01 16:10 & Strahl drop out and inversion\\
    5 & 2020-06-01 19:35 &  2020-06-01 21:35 & Flux rope and strahl inversion\\
    5 & 2020-06-02 06:50 &  2020-06-02 09:10 &Partial HPS crossing and MICCS\\
    5 & 2020-06-04 03:25 &  2020-06-04 06:05 & Partial HPS crossing with strahl drop out\\
    5 & 2020-06-07 11:10 &  2020-06-07 12:40 & Partial HPS crossing with strahl drop out\\
    5 & 2020-06-07 20:20 &  2020-06-07 21:10 & Partial HPS crossing with strahl drop out\\
    5 & 2020-06-08 00:00 &  2020-06-08 12:30 & HCS crossing\\
    5 & 2020-06-08 15:30 &  2020-06-09 01:40 & HCS crossing\\
    5 & 2020-06-12 01:00 &  2020-06-12 08:à0 & Flux rope or CME\\
    
    \end{tabular}
    \caption{Intervals discarded in our analysis where switchbacks are not defined}
    \label{tab:intervals}
\end{table*}

We display in the following set of figures \ref{figset: appendix_figures} the full window wavelet analyses of the encounters 1, 2, 4 and 5, both over time and space. They are displayed in a similar manner as in the text, with intervals of Table \ref{tab:intervals} indicated by greyed data in the timeseries, and blackened areas in the wavelet power spectrum. For $E_1$ we placed our analysis in-between the two CMEs (\cite{2020ApJS..246...63N}, \cite{2020ApJS..246...69K}) that occurred upon entry and exit of the encounter,leading to a shorter interval of analysis.

\figsetstart
\figsetnum{7}
\figsettitle{Wavelet analysis (temporal and spatial) for $E_1$, $E_2$,$E_4$,$E_5$}

\figsetgrpstart
\figsetgrpnum{7.1}
\figsetgrptitle{Encounter 1 - Temporal analysis}
\figsetplot{App_E1_time.png}
\figsetgrpnote{Encounter 1 - Temporal analysis}
\figsetgrpend

\figsetgrpstart
\figsetgrpnum{7.2}
\figsetgrptitle{Encounter 1 - Spatial analysis}
\figsetplot{App_E1_space.png}
\figsetgrpnote{Encounter 1 - Spatial analysis}
\figsetgrpend

\figsetgrpstart
\figsetgrpnum{7.3}
\figsetgrptitle{Encounter 2 - Temporal analysis}
\figsetplot{App_E2_time.png}
\figsetgrpnote{Encounter 2 - Temporal analysis}
\figsetgrpend

\figsetgrpstart
\figsetgrpnum{7.4}
\figsetgrptitle{Encounter 2 - Spatial analysis}
\figsetplot{App_E2_space.png}
\figsetgrpnote{Encounter 2 - Spatial analysis}
\figsetgrpend

\figsetgrpstart
\figsetgrpnum{7.5}
\figsetgrptitle{Encounter 4 - Temporal analysis}
\figsetplot{App_E4_time.png}
\figsetgrpnote{Encounter 4 - Temporal analysis}
\figsetgrpend

\figsetgrpstart
\figsetgrpnum{7.6}
\figsetgrptitle{Encounter 4 - Spatial analysis}
\figsetplot{App_E4_space.png}
\figsetgrpnote{Encounter 4 - Spatial analysis}
\figsetgrpend

\figsetgrpstart
\figsetgrpnum{7.7}
\figsetgrptitle{Encounter 5 - Temporal analysis}
\figsetplot{App_E5_time.png}
\figsetgrpnote{Encounter 5 - Temporal analysis}
\figsetgrpend

\figsetgrpstart
\figsetgrpnum{7.8}
\figsetgrptitle{Encounter 5 - Spatial analysis}
\figsetplot{App_E5_space.png}
\figsetgrpnote{Encounter 5 - Spatial analysis}
\figsetgrpend
\label{figset: appendix_figures}
\figsetend

\begin{figure*}
    \centering
    \includegraphics[width=1\textwidth]{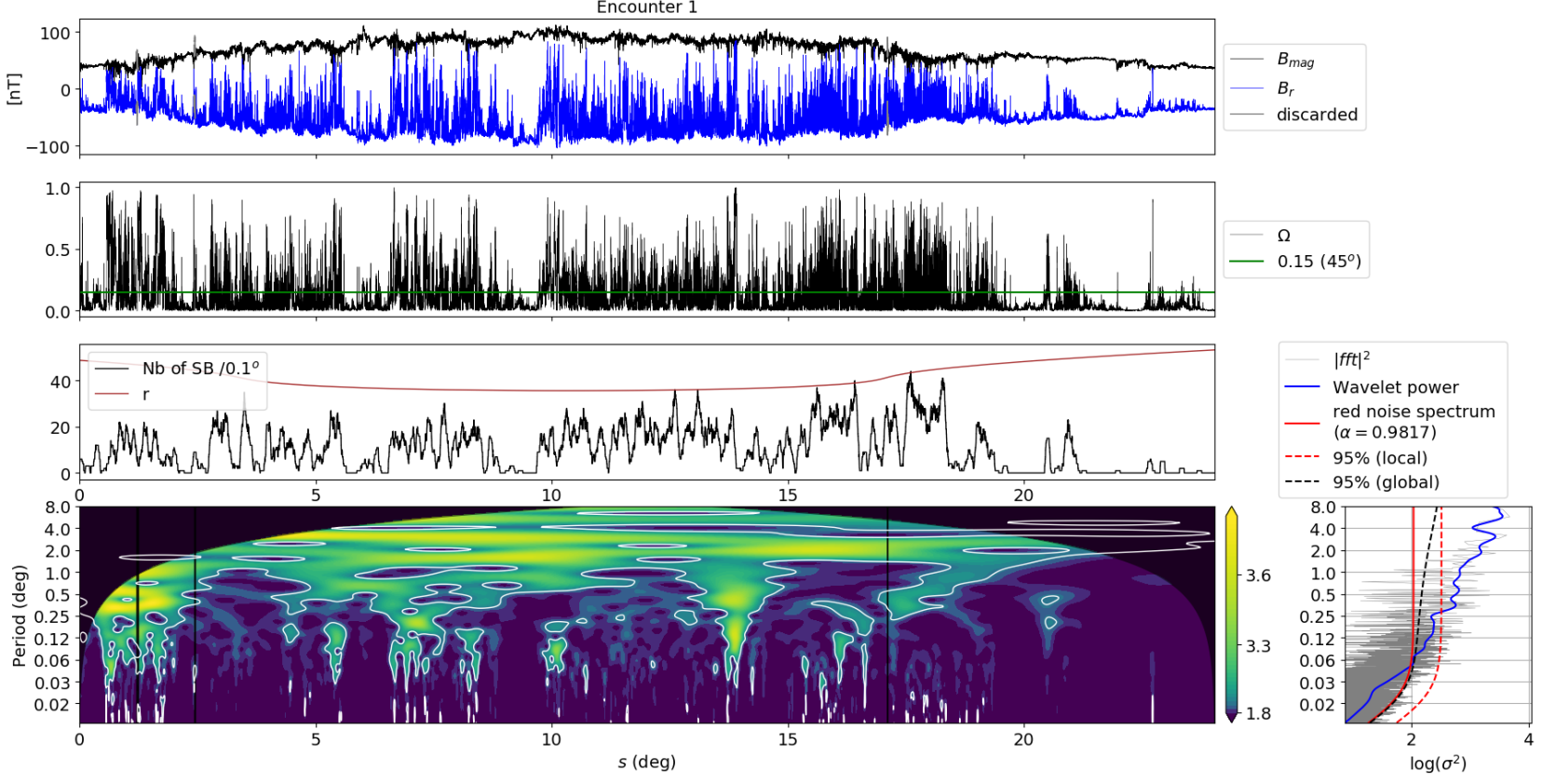}
    \caption{Encounter 1 - Spatial analysis. The complete figure set for encounters 1, 2, 4 and 5 (8 images) is available in the online journal. Data is displayed in a similar manner as in the text, with intervals of Table \ref{tab:intervals} indicated by greyed data in the timeseries, and blackened areas in the wavelet power spectrum.}
    \label{fig: app_5_s}
\end{figure*}

\end{document}